\documentclass{elsart}

\usepackage{graphicx}
\usepackage{amssymb}
\usepackage{soul}

\begin{document}
\begin{frontmatter}
\title{Radio-Detection Signature of High Energy Cosmic Rays by the CODALEMA Experiment}

\author[Nantes]{D. Ardouin}
\author[Nantes]{ A. Bell\'etoile}
\author[Nantes]{ D. Charrier}
\author[Nantes]{ R. Dallier}
\author[Nancay]{ L. Denis}
\author[Lal]{ P. Eschstruth}
\author[Nantes]{ T. Gousset}
\author[Nantes]{ F. Haddad}
\author[Nantes]{ J. Lamblin}
\author[Nantes]{ P. Lautridou \corauthref{ADD}}
\author[Paris]{ A. Lecacheux}
\author[Lal]{ D. Monnier-Ragaigne,}
\author[Nantes]{ A. Rahmani}
\author[Nantes]{ O. Ravel}
\corauth[ADD]{Corresponding Author: Tel.:+33-2-51858441; Fax:+33-2-51858479;
E-mail address: lautrido@in2p3.fr}

\address[Nantes]{SUBATECH, 4 rue Alfred Kastler, BP
  20722, F-44307 Nantes cedex 3}
\address[Nancay]{Observatoire de Paris - Station de radioastronomie,
  F-18330 Nan\c{c}ay}
\address[Paris]{LESIA, Observatoire de Paris, 5
  place Jules Janssen, F-92195 Meudon cedex}
\address[Lal]{LAL, Universit\'e Paris-Sud, B\^atiment 200, BP 34,
  F-91898 Orsay cedex}

\begin{abstract}
Taking advantage of recent technical progress which has overcome some
of the difficulties
encountered in the 1960's in the radio detection of extensive air
showers induced by ultra high energy cosmic rays (UHECR),  a new
experimental apparatus (CODALEMA) has been built and operated. We
will present the characteristics of this device and the analysis
techniques that have been developed for observing
electrical transients associated with cosmic rays.
We find a collection
of events for which both time and arrival direction coincidences between particle and radio signals
are observed. The counting rate corresponds to shower energies $\geq  5\times 10^{16}$~eV. 
The performance level which has been reached considerably enlarges the 
perspectives for studying UHECR events using radio detection.
\end{abstract}

\end{frontmatter}

%
%
\section{Introduction}

For almost 70 years, physicists and astronomers have studied cosmic
rays and gained a good knowledge of the flux as a function of energy
up to $10^{19}$~eV. However, in this energy range, the problem of the
origin and the nature of ultra-high energy cosmic rays (UHECR) is
unsolved and stands as one of the most challenging questions in
astroparticle physics. In order to collect the elusive events above
$10^{19}$~eV (which present an integrated flux of less than 1 event
per km$^2$ per steradian and per year) giant detectors are needed and
are presently being designed and built. Today, for studying the
highest energy extensive air showers (EAS), a leading role is played
by the Pierre Auger Observatory~\cite{auger} which uses a hybrid
detection system combining particles and fluorescence which inherently
have very different duty cycles.  An alternative method was suggested
long ago by Askar'yan~\cite{Ask62} consisting in the observation of
radio emission associated with the development of the shower. This
method is based on the coherent character of the radio emission and
deserves, in our opinion, serious reinvestigation due to many
potential advantages as compared to other methods. Besides expected
lower cost and sensitivity to the longitudinal shower development, it
primarily offers larger volume sensitivity and duty cycles which can
result in higher statistics, thus offering better possibilities to
discriminate between postulated scenarios of UHECR nature and
origin~\cite{sigl}.

Experimental investigations carried out in the 1960's proved the
existence of RF signals initiated by EAS (for a comprehensive review
see~\cite{Allan}). The evidence was obtained with systems consisting
either of a radio telescope or an antenna, sometimes a few antennas,
triggered by particle detectors. The bulk of the data was obtained in
narrow frequency bands (about 1 MHz), centered around various
frequencies in the 30--100~MHz range. The difficulties encountered were
numerous.  Among other
things, insufficient electronics performance and problems due to
atmospheric effects led to poor reproducibility of results, and this
area of research was abandoned in favour of direct particle~\cite{agasa}
and fluorescence~\cite{fly} detection from the ground.

But in recent years, the idea has once
again come to the forefront and new experiments have been undertaken
to detect radio pulses and measure their characteristics.
In one case, motivated by the perspective of large-scale cosmic ray
experiments, data were taken with a single antenna put into
coincidence with the CASA-MIA detector~\cite{casa-mia}.
In another case, in the perspective of the next generation of
low-frequency radio telescopes~\cite{Lopez,Lofar}, an array of antennas
was deployed to run in coincidence with the KASCADE air-shower
detector~\cite{cascade-grande}.

In this paper, we report on studies conducted with a radio air shower detector using a multi-antenna
array~\cite{rav04,dal03,dal04} set up at the Nan\c{c}ay radioastronomy
observatory.
From the earlier radio detection experiments, several
criteria were retained to provide optimal conditions for the detection
of transients: choice of a radio-quiet area, a well-understood radio
frequency (RF) electromagnetic environment, the use of several
antennas in coincidence and broadband frequency measurements. After a
discussion, in section~\ref{sec:motivation}, of the characteristics
predicted for the electric pulses generated by EAS,
section~\ref{sec:setup} will describe the CODALEMA (COsmic ray
Detection Array with Logarithmic ElectroMagnetic Antennas)
experiment together with the trigger definition in stand-alone mode. 
Observation of transient signals are discussed in
section~\ref{sec:transient} and the analysis techniques are outlined in~\ref{sec:signal} along with their
limitations. 
Particle detectors have been set up to provide a trigger 
on cosmic ray events. After a description of the analysis carried out in this mode of operation, 
section~\ref{sec:particules} shows that some of the observed transients 
originate from EAS. Conclusions and perspectives are given in the last section.

%
\section{Signal properties}
\label{sec:motivation}

An extensive air shower contains a huge number of electrons and
positrons (several billion at maximum for $10^{20}$~eV). At first
glance, the stochastic nature of such moving charge distributions
should produce incoherent fields. This is the case for the \v Cerenkov
radiation visible in the optical domain. It turns out, however, that
several physical effects systematically break the symmetry between the
electron and positron distributions inside the shower, leading to a
coherent field contribution in the RF domain. This is the case for the
so-called Askar'yan emission~\cite{Ask62} for which Compton scattering
and in-flight positron annihilation create an excess of electrons in
the shower front. In a recent experiment~\cite{slac} in which a GeV
photon beam was sent into a sand target several meters long, this
coherent emission was observed and its properties were found to be
consistent with the predictions.  Based on this phenomenon, radio
detection of cosmic ray induced showers in dense media is currently
under investigation~\cite{rice,moon}. The Askar'yan effect should also
exist for EAS, though it is likely to be dominated by the coherent
emission associated with another charge separation mechanism, charge
deflection in the Earth's magnetic field~\cite{Allan,kandl,huege}.

\subsection{Electric field characteristics}

In order to determine the feasibility of measuring electric fields
from high energy air showers, it is first necessary to have some
estimates for the maximal amplitude and the time scale(s) of the associated signal as well as their lateral
distribution at ground level.

\subsubsection{Around $10^{20}$~eV}

For a set of widely spaced antennas, most events will be at large
impact parameters. In this range, the time scale is primarily set by
the time development of the shower albeit with a certain amount of
Doppler contraction. Time differences between signals from different
individual charges at a given shower age are generally small on the
above time scale. Consequently, a good assumption is provided with a
point-like charge approximation. The interest of such a model is that
electric field behavior is easy to obtain with standard textbook
formulas~\cite{Allan,jackson}.

The easiest field contribution to evaluate is that due to charge
excess~\cite{Ask62}. In principle, the excess is a function of
particle energy, but this effect can be ignored in a first
estimate. Thus, $q(t)=-\varepsilon N(t) e$, with $N(t)$ the number of
electrons at a given time and $\varepsilon$, the fractional excess. At
an observation point $A$ and at time $t$, the electric field for a
moving point-like charge is given by
\begin{eqnarray}
\mathbf{E}(t,A)&=&\frac{1}{4\pi\epsilon}\sum_{t'}(1-v^2/c^2)q
\frac{\mathbf{R}-R\mathbf{v}/c}{|R-\mathbf{R}\cdot\mathbf{v}/c|^3}
\nonumber\\
\label{eq:electric-field}
&&+\frac{1}{4\pi\epsilon c}\sum_{t'}\dot{q}
\frac{\mathbf{R}-R\mathbf{v}/c}{(R-\mathbf{R}\cdot\mathbf{v}/c)
|R-\mathbf{R}\cdot\mathbf{v}/c|}
\end{eqnarray}
where $\mathbf{R}=\overrightarrow{QA}$ is the vector between the
charge position at time $t'=t-R/c$ and the observation point and
$\mathbf{v}$ is the charge velocity.

\begin{figure}
\begin{center}
\vspace*{8pt}
\includegraphics[width=10cm]{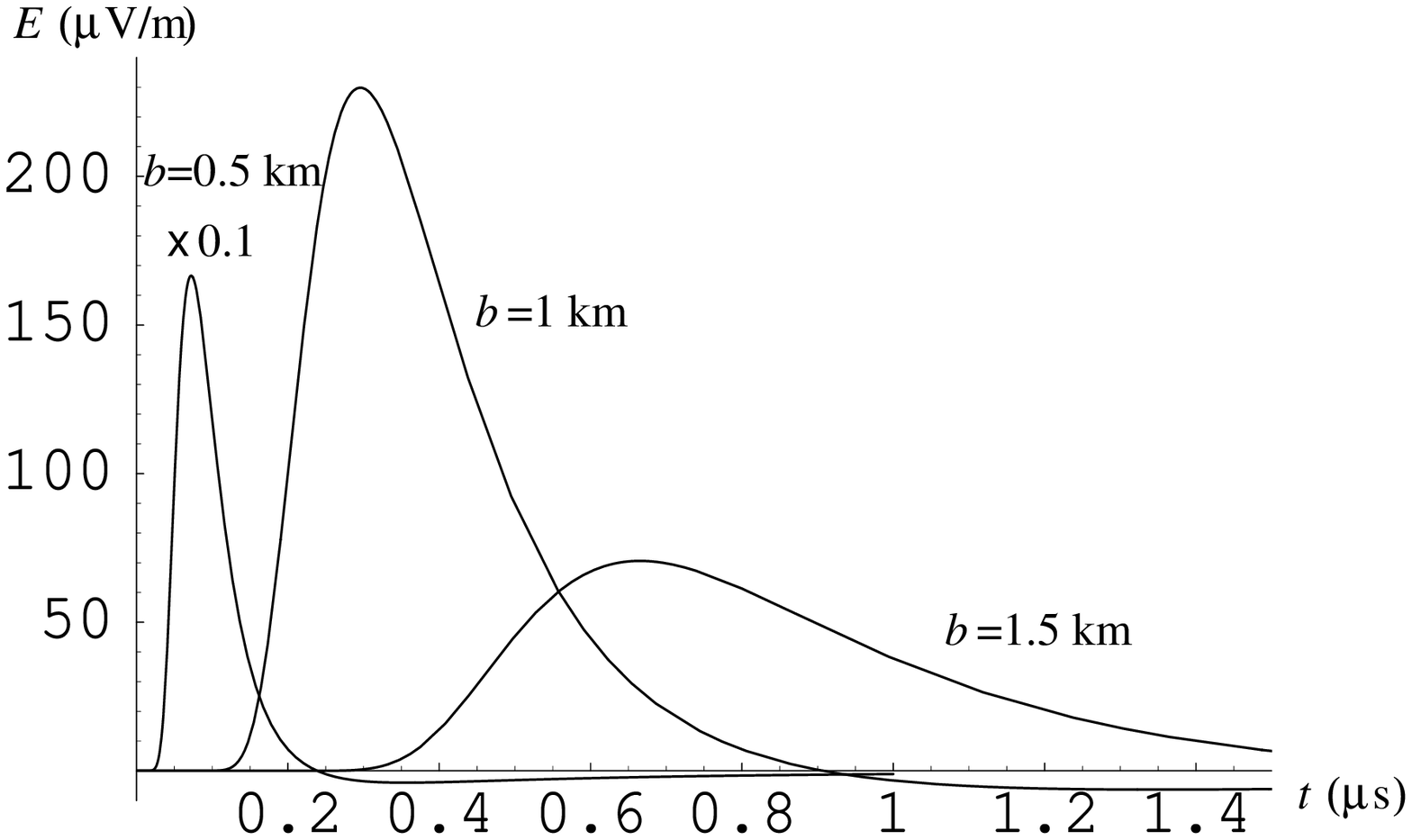}\\
\includegraphics[width=10cm]{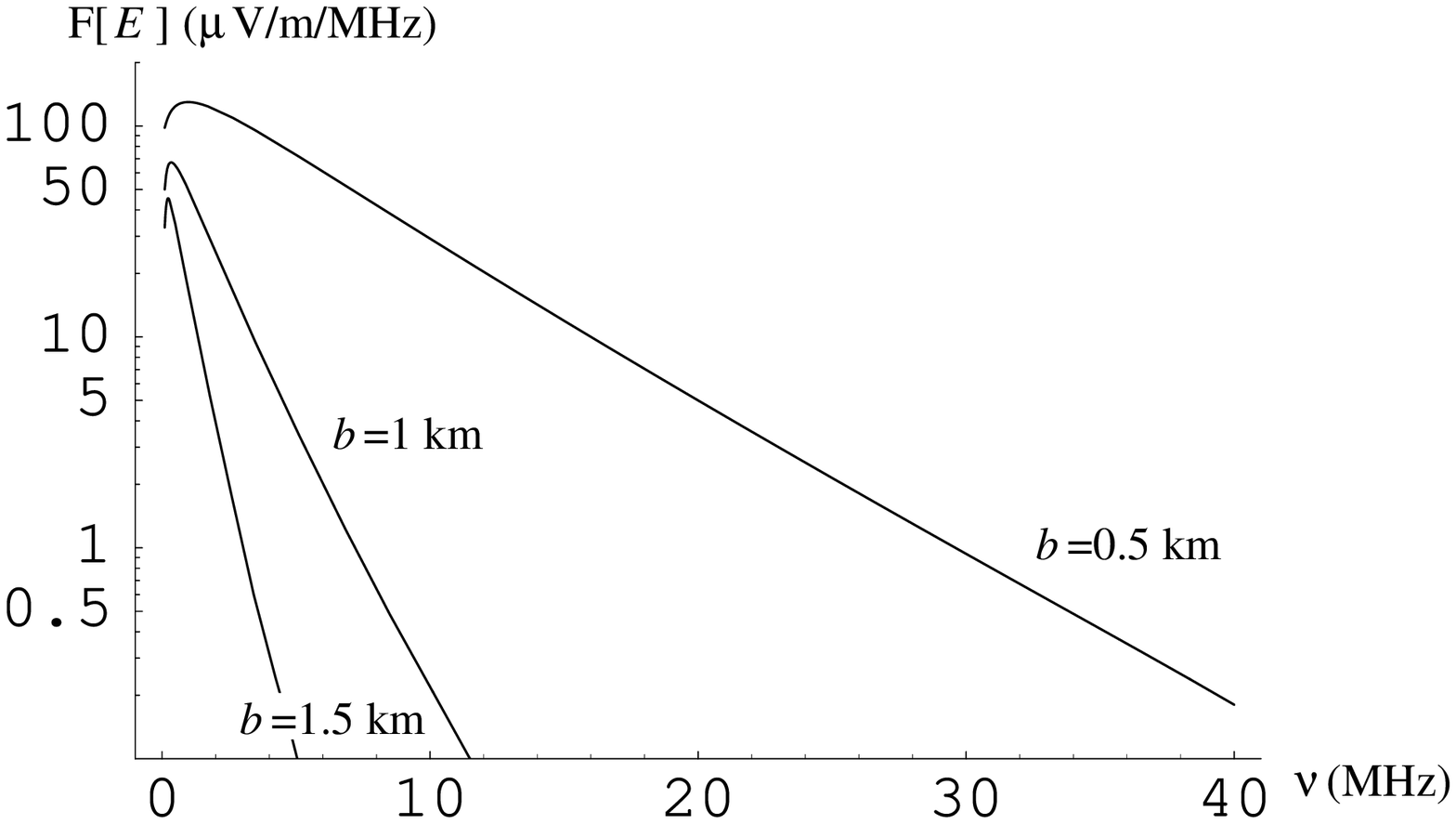}
\end{center}
\caption{Top panel: electric field as a function of time for various
impact parameters at sea level for a $10^{20}$~eV vertical shower
with $\varepsilon=10\%$ and $Xmax=1700$~m. Bottom panel: corresponding magnitude
coefficient distributions obtained from the Fourier transforms.}
\label{fig:excess}
\vspace*{12pt}
\end{figure}

In the upper part of Fig.~\ref{fig:excess}, the horizontal component
of the electric field is plotted as a function of time for a
$10^{20}$~eV vertical shower ($\varepsilon=10\%$, $Xmax=1700$~m)
with impact parameters $b=0.5$, 1~km and 1.5~km . At $b=1$~km the
field reaches 230~$\mu$V$/$m.
Such sizable values can be obtained only for huge
cosmic ray energies. This points also to the fact that long distance
detection is only possible for extremely high energies. The lower part
of Fig.~\ref{fig:excess} presents the associated Fourier spectra
indicating that very broadband antennas must be used in order to
maintain sufficient sensitivity in the MHz range.  The coherence,
which is essential in order to reach such high values, is lost at
high frequency where the structure of the shower becomes resolved.
These limitations are moderated somewhat in the case of inclined
shower configurations~\cite{gousset} and by the fact that for most
shower orientations the dominant field contribution is likely to come
from charge deflection in the Earth's magnetic
field~\cite{Allan,kandl,huege}.

An inspection of Eq.~(\ref{eq:electric-field}) shows that the electric
field waveform gives an image of the longitudinal development of the
shower. In this respect, the information obtained by the radio
technique is comparable to that of fluorescence detectors, though this
radio image is distorted by the Doppler effect and by the
ultra-relativistic nature of the emitting charges.

\subsubsection{Around $10^{17}$~eV}
\label{sec:1017}

Because the rate of $10^{20}$~eV cosmic rays is very small, an 
 experiment would need a setup covering a large surface in order to collect enough statistics.
 Consequently it has been decided to work at much lower energy. 
Because the energy radiated through coherent emission is expected to be
proportional to the square of the number of charges within the shower
(i.e. proportional to the square of the energy of the primary), signals
from $10^{17}$~eV air showers will be appreciably smaller than at $10^{20}$~eV. At large distance from the shower core,
this effect will be strengthened. Therefore,
the contribution to the rate from EAS will be dominated by those whose
cores fall within (or close to) the area delimited by the antennas and
with a small number of antennas, interesting events are likely to be
those that are detected by the majority, if not all, of them.
Taking into account the inter-antenna distance of our setup (see
\ref{sec:setup}), an estimate of the electric field must be made for
impact parameters in the 100 meter range. In this domain of impact parameters,
no simple expression for the electric field
can be given. Recent modelizations exist regarding this
question~\cite{huege,suprun}, but here we will use estimates as
provided in Refs.~\cite{Allan} and \cite{casa-mia}.

The pulse waveform can be written
\begin{equation}\label{eq:waveform}
f(t)=\theta(t) A e^2 \left(\frac{t}{2\tau_1}\right)^2
\left(e^{-t/\tau_1}-(\tau_1/\tau_2)^3 e^{-t/\tau_2}\right),
\end{equation}
where $\theta(t)=1$ for $t>0$ and $0$ for $t<0$. This expression is
derived from various general considerations concerning the shower's
electric field. In particular, the second exponential represents a
negative-amplitude component such that the signal has no DC
component. Its duration is, at least for a vertical shower, much
longer than the positive-amplitude component ($\tau_2\gg\tau_1$) and
is, therefore, of minor importance for the present discussion. For
$\tau_2\gg\tau_1$ the relevant parameters are thus $\tau_1$ which
fixes the time scale of the positive-amplitude part (the maximum is
reached at time $t=2\tau_1$) and $A$ the maximum amplitude.

\begin{figure}
\begin{center}
  \vspace*{8pt}
\hbox{\includegraphics[width=7.5cm]{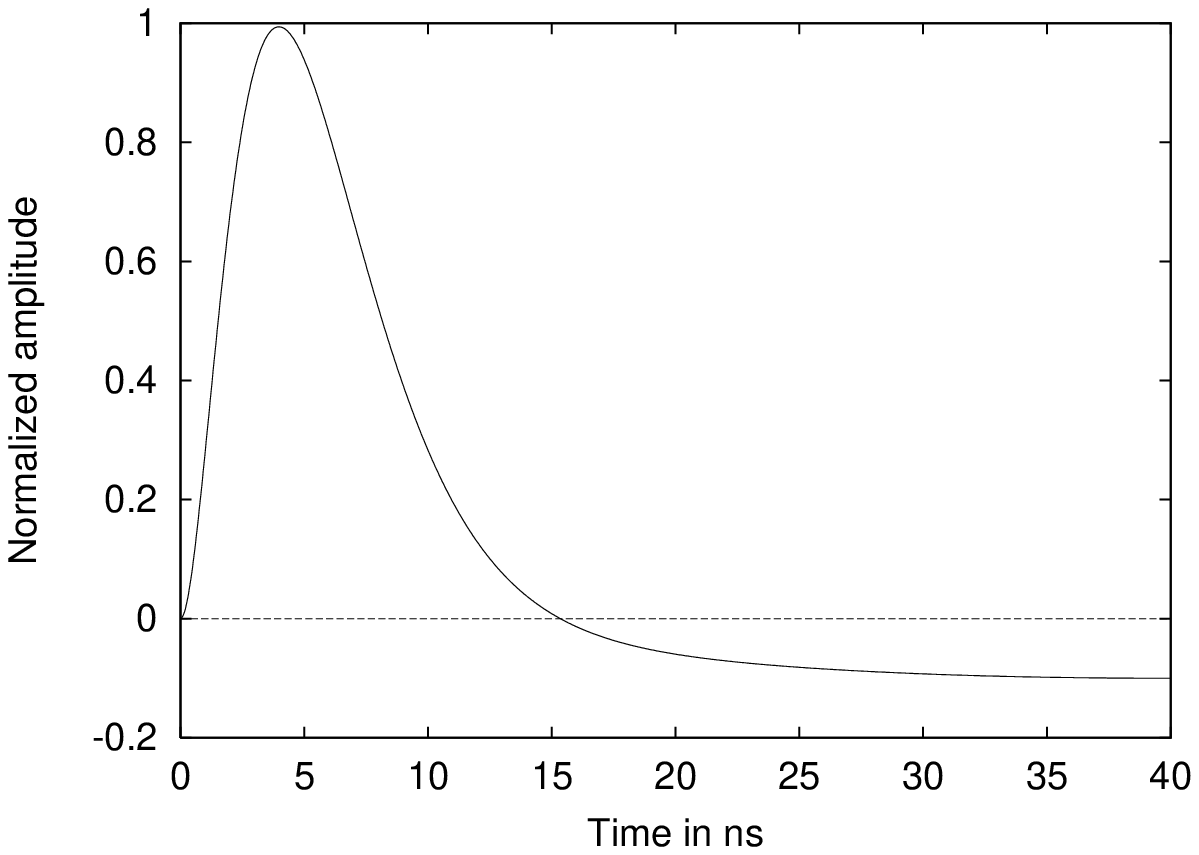}
\includegraphics[width=7.5cm]{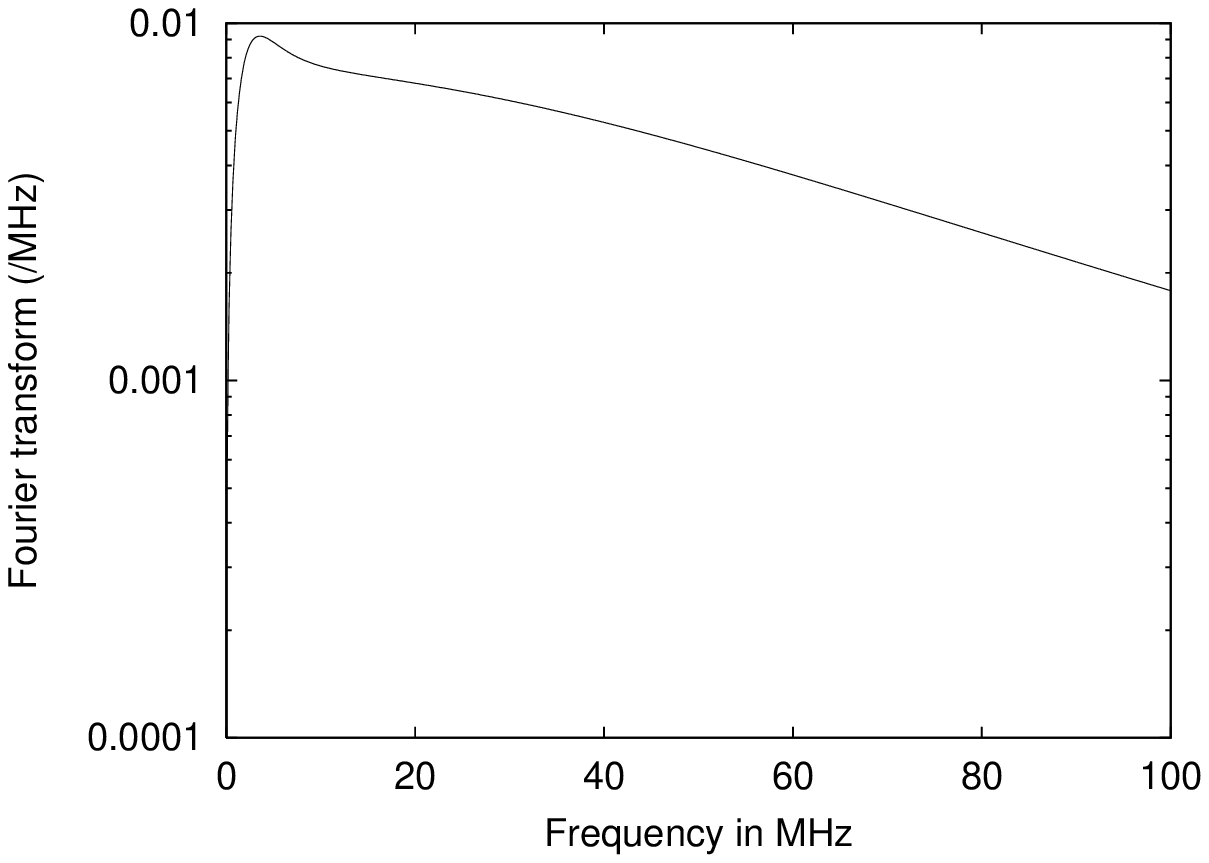}}
\end{center}
\caption{Pulse waveform (left) and corresponding Fourier transform $|S(\nu)|$ (right) obtained from Eq.~(\ref{eq:waveform}) using the parameters $\tau_1=2$~ns, $A=1$ and $\tau_2=20$~ns.}
\label{fig:waveform}
  \vspace*{12pt}
\end{figure}

For a $10^{17}$~eV vertical shower at small impact parameters,
Allan~\cite{Allan} estimates the pulse duration to be of the order
of 10~ns and the electric field amplitude $A\sim 150$~$\mu$V$/$m. The
pulse waveform from Eq.~(\ref{eq:waveform}) with $\tau_1=2$~ns, $A=1$
and $\tau_2=10\times\tau_1$, and the corresponding modulus of the
Fourier transform are shown in Fig.~\ref{fig:waveform}.

\subsection{Filtering and triggering}
\label{subsec:filtering}

\begin{figure}
\begin{center}
  \vspace*{8pt}
\hbox{\includegraphics[width=7.5cm]{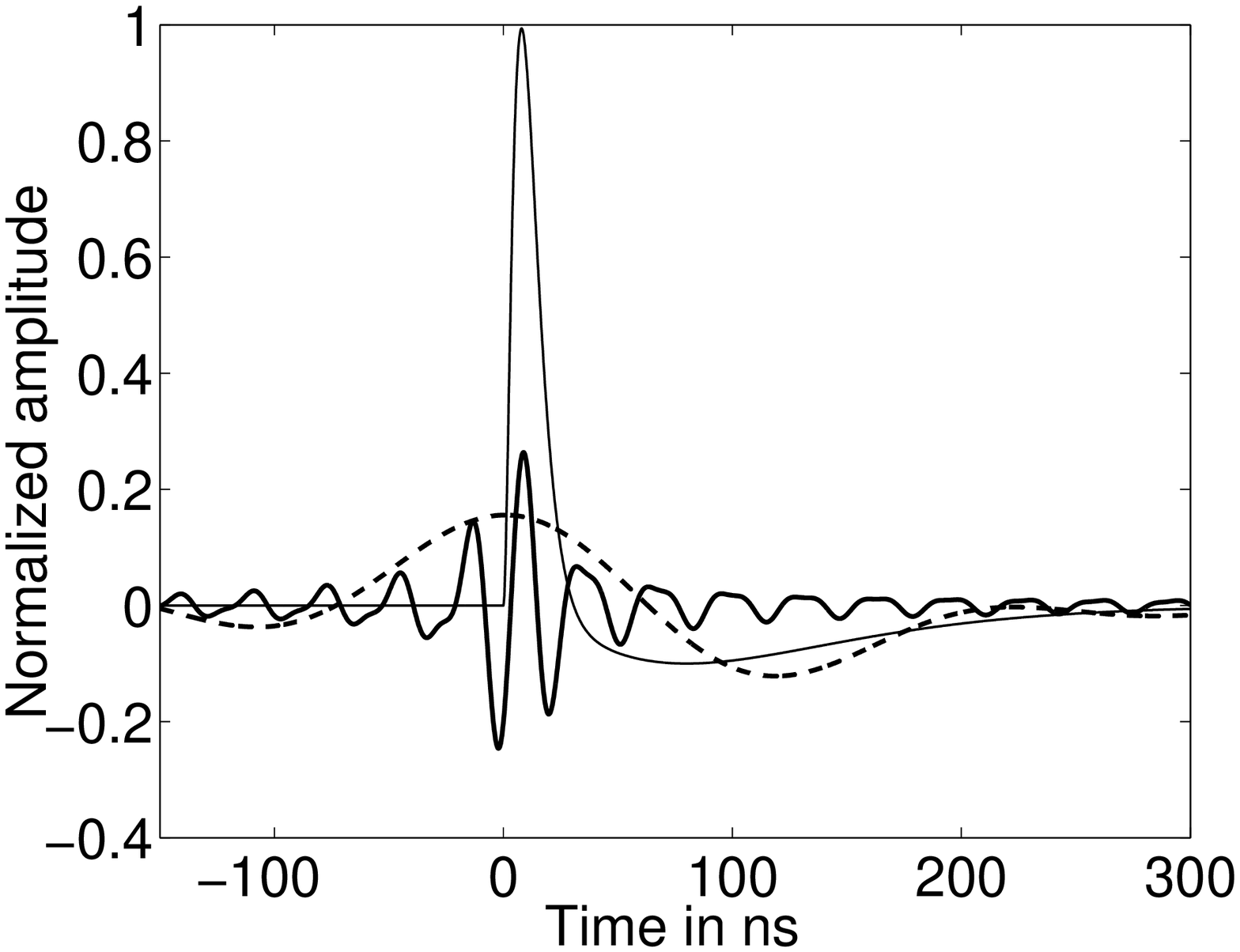}
\includegraphics[width=7.5cm]{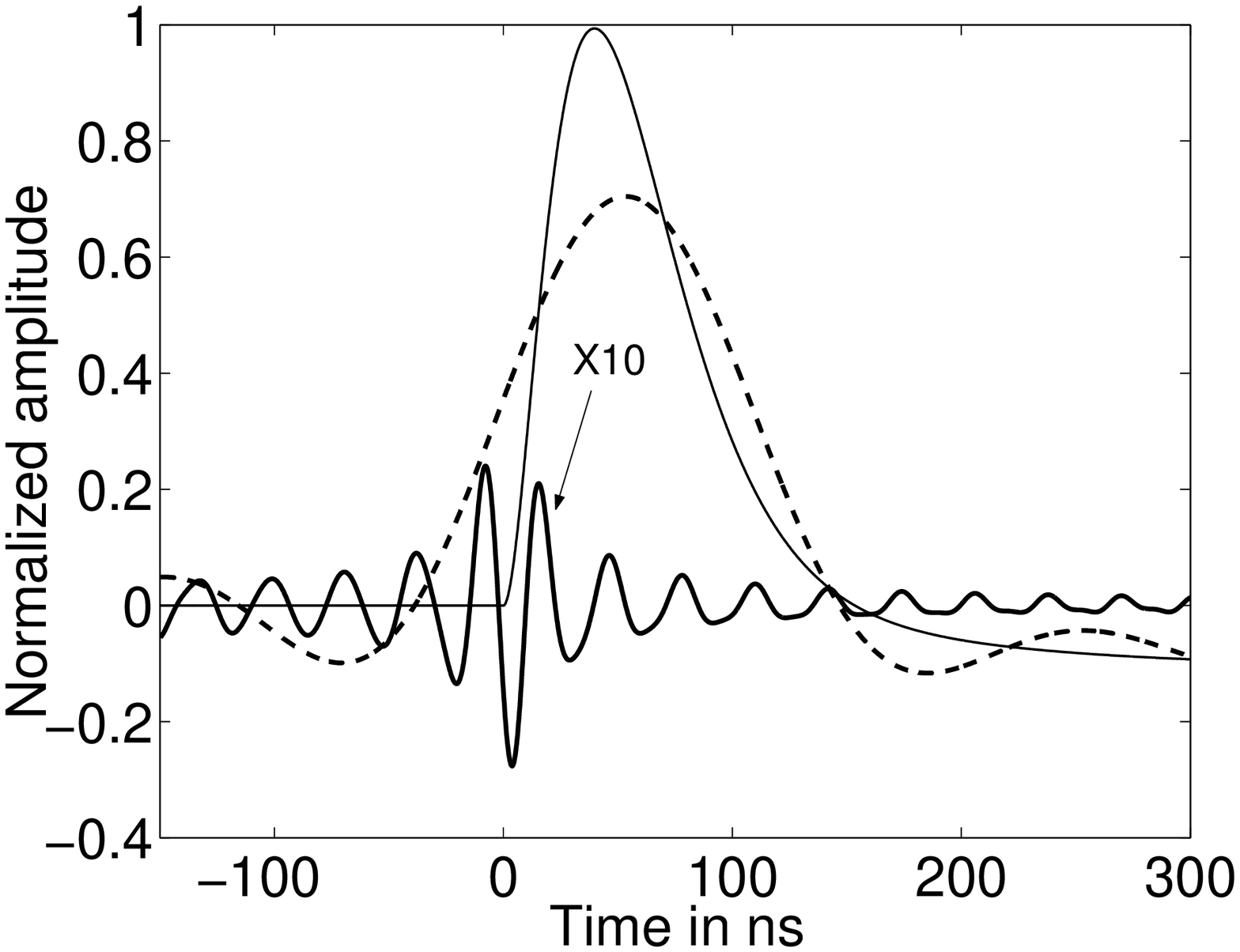}}
\end{center}
\caption{Raw pulses (thin line) and their band-pass filtered
  counterparts for bandwidths of 30--60~MHz (thick full line) and 1--5~MHz
  (dashed line).}
\label{fig:filtered_signal}
  \vspace*{12pt}
\end{figure}

From the experimental point of view, the electrical signal will be
distorted when passing through the detector chain consisting of
elements such as antennas, cables and preamplifiers.  Insertion losses
and filtering effects will decrease the signal amplitude and strongly
modify the initial waveform.  The effect of band-pass filtering is
shown in Fig.~\ref{fig:filtered_signal} for a short duration pulse
such as that of Fig.~\ref{fig:waveform} as well as a much longer one.

Whatever the duration of the pulse may be, the high frequency
component (30--60~MHz) is correlated with the rise time
of the signal and is suitable for trigger purposes. In addition,
depending on its duration, analysis of the low frequency part of the
signal (1--5~MHz) can lead to a direct estimate of the signal
amplitude.

The capability to detect the radio signal strongly depends on the
choice of the frequency band. Nevertheless, the simultaneous presence
of the time-limited oscillating signals in two widely separated
frequency bands enables us to distinguish broadcast emissions, which
are quasi-monochromatic, from broadband signals such as the ones
expected from EAS.

%
%
\section{The experimental set-up}
\label{sec:setup}

One of the originalities of CODALEMA lies in the fact that the system
can be self-triggered using a dedicated antenna, as opposed to the other
experiments mentioned above where particle detectors provide the
triggering. However, for such an experiment, it is then
necessary to become acquainted with the characteristics of the
transient RF sky, which, in addition, is almost completely
unexplored for time scales ranging from nanoseconds to
microseconds. The decision was made to tackle the problem
with a small number of antennas arranged in a ``self-triggered
all-radio'' system, intrinsically suited to measuring the RF transient
rate. We expected this task to be
facilitated by the progress accomplished in electronics and data
processing since the 1960's. In the meantime however, the sky has
generally become much noisier than at the time of the pioneering work.
Thus the only condition imposed for recording data is the arrival of a
suitable RF transient on a trigger antenna, provided that the dead time
due to the data recording (i.e. the trigger rate), stays at an
acceptable level. Under these conditions, the duty cycle stays close
to 100\% and the setup should be well suited for registering possible incoming EAS signals.

An alternative strategy would have been to design a system where
antenna signals are sampled when a particle trigger occurs. In fact
this method, which has its own physical limitations, is now being
used~\cite{ardouin} and will be described in~\ref{sec:particules}.
However, for such designs the sky transient
background is inaccessible because the overall observation time
remains very small: as an example the time window ($50~\mu$s) which is
used in~\cite{casa-mia} in association with the low particle-trigger rate,
leads to an integrated observation time of only a small fraction of
a second (50~ms), for a typical day-long run.

\subsection{General layout}

\begin{figure}
\centering
  \vspace*{8pt}
\includegraphics[width=14cm]{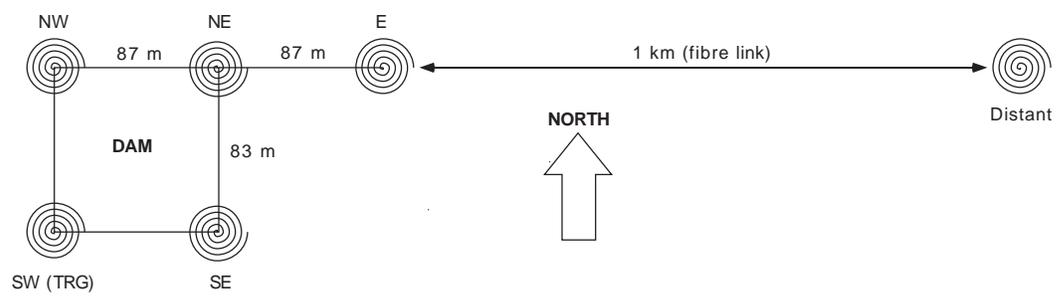}	
\caption{The CODALEMA setup, first phase.}
\label{fig:setup}
  \vspace*{12pt}
\end{figure}

\begin{figure}
\centering
  \vspace*{8pt}
\includegraphics[width=10cm]{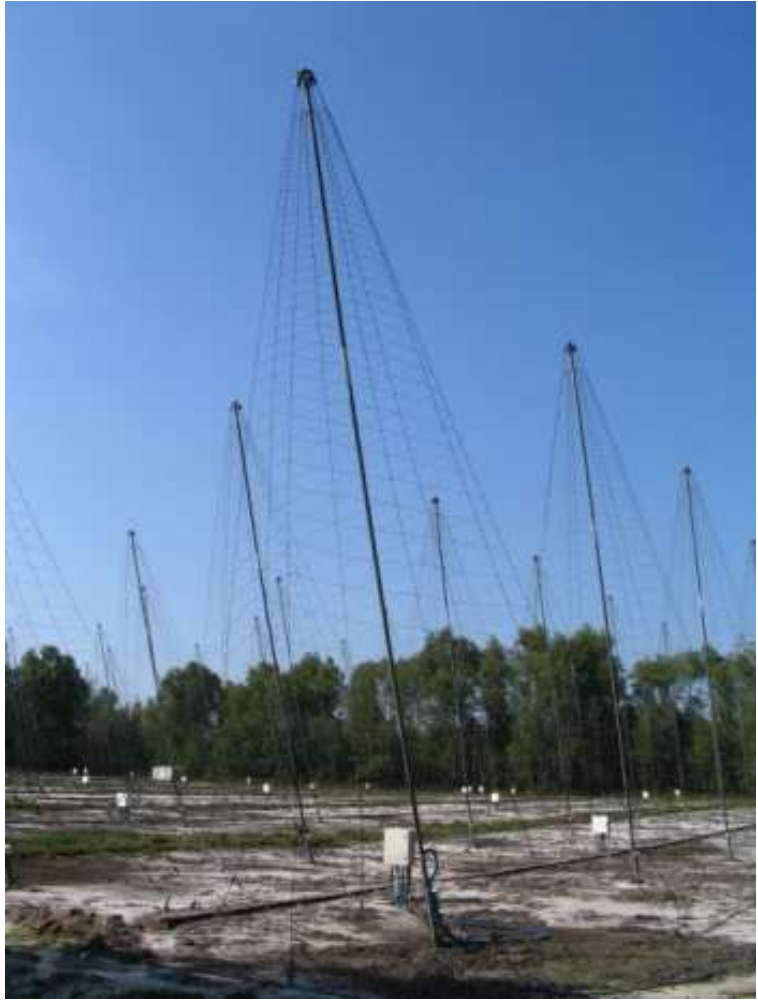}	
\caption{Photograph of the logarithmic antennas of the Nan\c{c}ay DeCAmetric Array. The CODALEMA experiment
uses some of these antennas.}
\label{fig:antenne_DAM}
  \vspace*{12pt}
\end{figure}

Running from March 2003, the CODALEMA experiment in its first phase used 6 of the 144
log-periodic antennas  of the DecAMetric array (DAM)(see figure \ref{fig:antenne_DAM}), 
an instrument
dedicated to the observation of the sun and Jupiter based at the
Nan\c{c}ay Radio Observatory~\cite{RDN}.  Having 1--100~MHz frequency
bands, these antennas are well adapted to our sensitivity and
frequency requirements. The antenna locations are shown in
Fig.~\ref{fig:setup}.

Four of the antennas, namely NE, SE, SW and NW, are located at the
corners of the DAM array. This layout was chosen in order to have
distances between detectors as large as 120 m and to minimize the
cable lengths. Two additional antennas, namely E and Distant, were
setup to the east of the DAM at respectively 87~m and 0.8~km providing
longer inter-antenna distances. From the information on the electric
field strength far from the central array, it is possible to identify
and eliminate broadcast signals and strong interference phenomena
(storms, etc.), expected to irradiate widely-separated antennas with
more or less uniform power.

 RF signal amplification(1--200 MHz, gain 35 dB) is performed using commonly available low-noise electronic
devices which have a negligible impact on the overall noise of the electronic chain.
The five grouped antennas are linked via 150~m coaxial cables (RG214U) to LeCroy
digital oscilloscopes (8 bit ADC, 500~MHz frequency sampling, 10~${\mu}$s recording
time). The Distant antenna requires a different signal transmission
technique, an optical fiber link approximately 1~km long which
introduces significant attenuation (a factor of 10), a 5.5~${\mu}$s
delay as well as a low frequency cutoff at 10 MHz from the
analog/optical transceiver. Special attention is paid to the shielding
both of the electronics itself and the acquisition room in order to
minimize interference coming from data-taking activities.

\subsection{Trigger definition}
\label{sec:trigger_definition}

\begin{figure}
\begin{center}
  \vspace*{8pt}
\includegraphics[width=12cm]{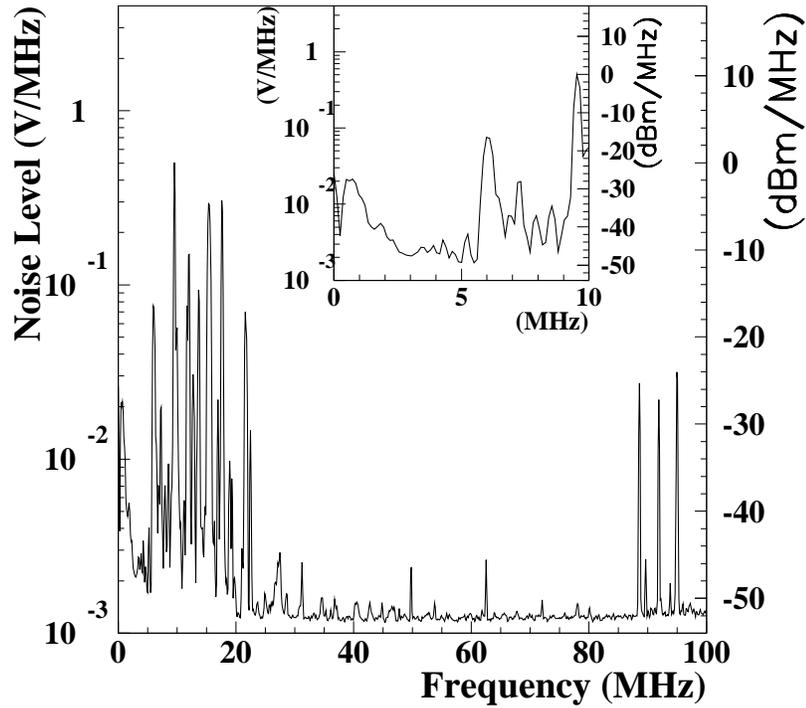}
\caption{Average Fourier transform $|S(\nu)|$ of the background
  voltage. The correspondence with the power spectral density in a
  $50~\Omega$ load impedance is shown on the right scale. This
  spectrum has been obtained by averaging power spectral densities of
  900 random events recorded with a sampling frequency of 500~MHz during 10~${\mu}$s every 10~s.
  A zoom covering the 0--10~MHz band is presented in the inset.}
\label{fig:mean_spectre_radio_Nancay}
\end{center}
  \vspace*{12pt}
\end{figure}

A trigger with minimum bias was chosen in order to select potentially
interesting events.  It is sensitive to the unusual frequency
contributions which come from transient signals. The amplitudes of
these contributions are compared to the normal sky background level,
whose frequency content has been precisely measured on site with one
of the antennas using the complete acquisition chain. The power
spectral density from this
measurement is shown in Fig.~\ref{fig:mean_spectre_radio_Nancay}.
Above 90~MHz, peaks associated with FM radio signals are clearly
observed. Between 20~MHz and 90~MHz, a rather quiet band is found
reaching $10^{-3}$ V/MHz. Below 20 MHz, numerous transmitter spectral
lines result, with our spectral resolution, in a quasi continuous
contribution far above the minimum. Nevertheless, a quieter band
($\simeq 2.10^{-3}$ V/MHz) can be found between 1~MHz and 5~MHz as
shown in the inset of Fig.~\ref{fig:mean_spectre_radio_Nancay}.
 Evaluations of the noise characteristics have also been done in several other
places showing that the spectral profile depends strongly on the experimental
location.  For example, the CASA-MIA collaboration in the Utah desert observed interference
 at 55 MHz from a TV transmitter over
100 km away.  However, concerning the Pierre Auger Observatory site \cite{GAP}
  (Malargüe, Argentina), we recently monitored the R.F. sky there and found that
in the frequency windows of interest it is as clean as the Nan\c{c}ay site.
Thus the main constraints come from the proximity of human activities, 
and a significant advantage of places such as Nan\c{c}ay and the Auger site is to present
very quiet environments.

As explained in section~\ref{sec:motivation}, the frequency
distribution of the expected transients shows a wide band contribution.
A signal reaching the setup will add its frequency
contributions to the noise frequency distribution. Due to the
structure of the noise spectrum, the resulting effect will be more
easily seen in the two quiet bands discussed above, with the better
signal to noise ratio obtained for the 20--90~MHz range.
Moreover, for timing considerations, it is also important to find a quantity which 
will determine the arrival time of the signal. Based on the information presented in Fig.~\ref{fig:filtered_signal},
 the maximum absolute value of the oscillating high frequency filtered signal, which is
strongly correlated to
the leading edge of the signal, was chosen. In addition, because the associated 
error is then related to the pseudo period of the oscillations, the higher the filtering 
frequency, the lower the
uncertainty is.  Taking into account these considerations, it was thus
decided to use only part of the 20--90~MHz range, namely
33--65~MHz, for the trigger antenna. The quiet 1--5~MHz band may then
be used to quantify the low frequency contribution of any wide band
signal.

To generate the trigger, the signal of the antenna chosen for this
purpose is sent through an analog band-pass filter (33--65~MHz) to its
corresponding oscilloscope channel. An internal leading-edge discriminator
condition applied to this channel sets the amplitude threshold used to initiate
data recording. For each trigger sequence, the data from all the antennas are stored on
disk for off-line analysis.

\subsection{Field sensitivity}
\label{sec:digit}

\begin{figure}
\centering
  \vspace*{8pt}
\includegraphics[width=10.cm]{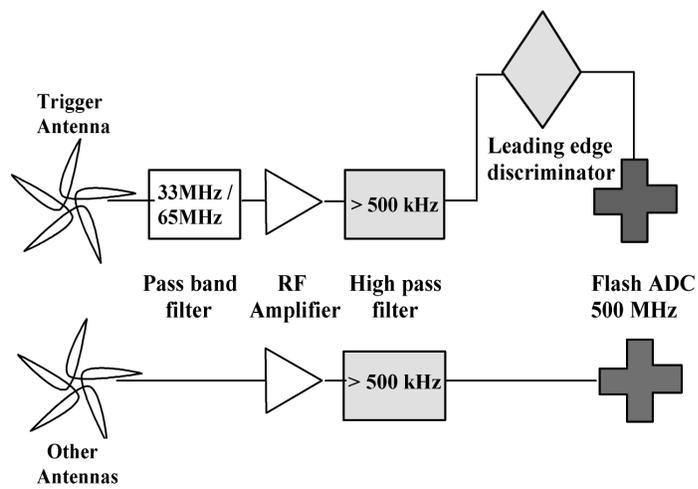}
\caption{CODALEMA electronics chain. }
\label{fig:electronics}
  \vspace*{12pt}
\end{figure}

The electronics setup used is shown in Fig.~\ref{fig:electronics}.
Following RF amplification, all the signals go through a high-pass
filter ($\geq$ 500 kHz) to remove the contribution of an AM
transmitter (164~kHz) located 22 km south of Nan\c{c}ay.
 The signal is digitized using a 8-bit ADC at 
a sampling frequency of 500~MHz with a 10~${\mu}$s recording
time.
Broadband
waveforms are recorded for all the antennas but the trigger, for which
the 33--65~MHz band-pass filter is present as indicated above.
 For the
other antennas, two different broadband configurations have been used
during the experiment. The antenna sensitivity rapidly decreases above
100~MHz, thus for analysis purposes, the ``full band'' configuration
has been restricted to 1--100~MHz and has been used to search for low
frequency counterparts of the transients. The ``restricted band''
configuration (analog filtering between 24 and 82~MHz) presents a
better signal-to-noise ratio and has been devoted to the search of
weak transients. Table~\ref{tab: sensitivity} indicates the
sensitivity reached by the apparatus for the different configurations.

This trigger sensitivity has to be compared with the expected value
for a $10^{17}$~eV vertical shower with a small impact parameter. The
values chosen for $A$ and $\tau_1$ are respectively 150~$\mu$V$/$m and
2~ns (see section~\ref{sec:1017}).
At this energy, in the frequency band $\nu_0\pm\Delta\nu/2=50\pm
15$~MHz, the expected field should give a peak amplitude
\[
E_{\mathrm{pk}}=2|S(\nu_0)|\Delta\nu=55\ \mu\mathrm{V/m}
\]
where $S(\nu_0)$ is the Fourier transform at frequency $\nu_0$ (see
Fig.~\ref{fig:waveform}).

This gives a voltage amplitude on the terminal resistance
$R=50~\Omega$~ \cite{casa-mia,kraus}
\[
V_{\mathrm{pk}}=E_{\mathrm{pk}}\frac{c}{2\nu_0}
\sqrt{\frac{RG}{\pi Z_0}}=2\ \mathrm{mV}
\]
where $Z_0=377~\Omega$ and $G$ is the overall gain from the antenna to
the terminal resistance including the antenna gain ($7$~dB), the
amplifier gain ($35$~dB) and the attenuation of the line ($-7$~dB).
These values clearly show that the electronic sensitivities in this
frequency band are adapted to the requirements for detecting EAS.

Concerning the trigger, it had been found that the background sky
noise at Nan\c{c}ay, referred to as $\sigma_{sky}$ in this paper, has
a standard deviation of 0.5 mV within the frequency band of
interest. In order to exclude most of the noise, we chose to set the
leading edge discriminator threshold to a value greater than
$4\,\sigma_{sky}=2$~mV which, as shown above, sets the lower energy
limit for detection to around $10^{17}$~eV.

This setup was operated on a regular basis between March 2003 and June
2004.  During that time $3.2\times10^3$~hours of running time
corresponding to $5.1\times10^4$ triggers were accumulated, mostly
devoted to setup and optimization of the device (electronics,
trigger antenna position, trigger threshold,... ).
However, during the last
six months of this period, data were taken for the sole purpose of
physics observations. The ``Full band'' configuration (see
table~\ref{tab: sensitivity}) was used for most of the runs with a
trigger threshold ranging from 4 to 10~mV. These runs were devoted to
the search for strong electric field pulses in the microsecond range
coming either from the sun, Jupiter or storms.  These data are
currently under analysis.  In June 2004, runs devoted to searching for
radio emission from EAS were carried out (see next section).  Filters were installed
on all except the trigger antenna in order to increase the sensitivity
(``restricted band'' configuration) and the trigger threshold was set to
2~mV.

\section{Transient signals}
\label{sec:transient}

\subsection{Trigger rates}

\begin{figure}
\centering
  \vspace*{8pt}
\includegraphics[width=12cm]{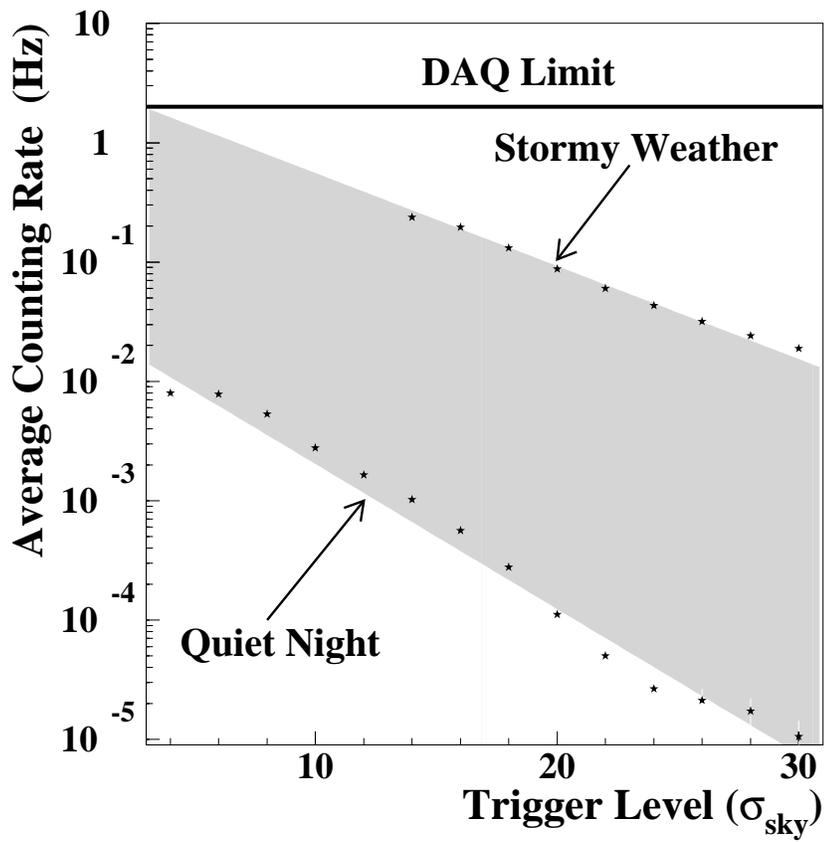}
\caption{Evolution of the counting rate as a function of the discriminator
threshold.}
\label{fig:triggerlevel}
  \vspace*{12pt}
\end{figure}

\begin{figure}
\centering
\vspace*{8pt}
\includegraphics[width=12cm]{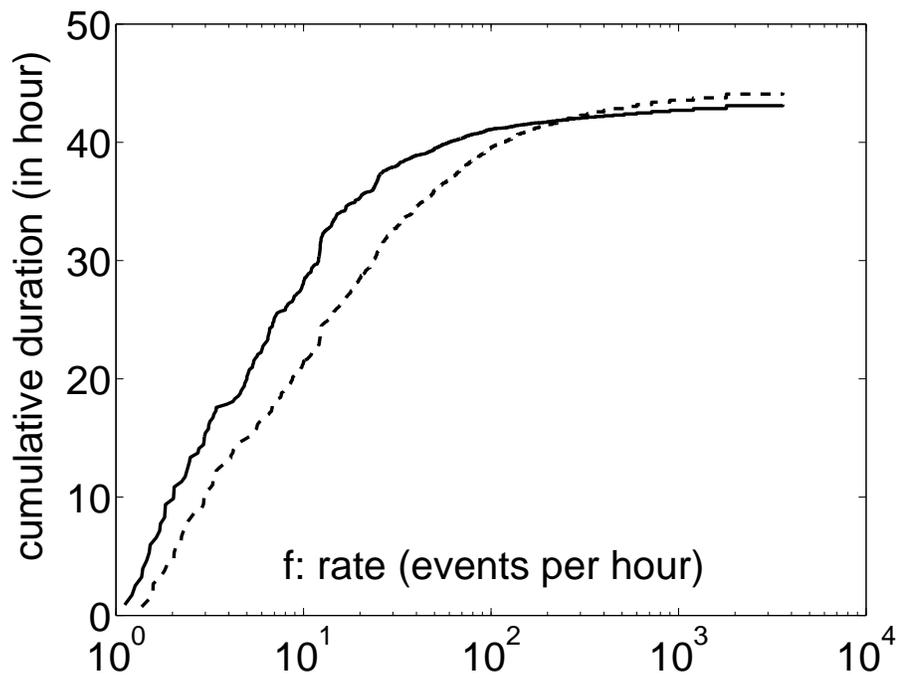}
\caption{Cumulative running time at a rate smaller than $f$ as a
function of the instantaneous rate $f$, for events with
several antennas in coincidence (full line) and for all
registered events (dashed line)}
\vspace*{12pt}
\label{fig:cumul}
\end{figure}

In Fig.~\ref{fig:triggerlevel} the average counting rate at Nan\c{c}ay
is presented as a function of the trigger level, expressed in units of
$\sigma_{sky}$. The counting rate evolves greatly with human
activities in the vicinity of the station as well as with the weather
conditions. The shaded area corresponds to the measured counting
rates. The lower limit was obtained during quiet night runs whereas
the upper one corresponds to stormy weather. Even then, the
acquisition system's maximum rate of two events per second is rarely
reached.  During the quietest nights very low counting rates were
obtained, making possible the detection of electric fields as low as
50~$\mu$V$/$m. Among the events recorded, only a few correspond to
an electromagnetic
(EM) wave crossing the apparatus. In order to identify these events,
we have compared two event configurations: one using all
the recorded events, the second presenting coincidences between the TRG, SE
and NW antennas. The procedure used to process the signals from the
antennas is presented in section~\ref{sec:signal}. The cumulative
running time obtained with this criterion is shown in
Fig.~\ref{fig:cumul} as a function of the instantaneous rate $f$ which
is defined as $f=\frac{1}{\Delta t}$, where $\Delta t$ is the time
between two consecutive events. The figure includes a total of 44 hours
of data-taking time at a threshold level of 4~$\sigma_{sky}$.

As can be seen, for both event topologies, the rate was smaller
than 10~events per hour for a
total of about 30 hours. These values can be compared with the
background rates observed in Ref.~\cite{casa-mia}, where the first
level selection gives an average of more than one transient every
50~$\mu$s. Though occasional counting rates greater than one or two~Hz
cannot be excluded with the present setup, Fig.~\ref{fig:cumul} shows
that there are large radio-quiet time intervals. The conclusion is
therefore that the sky at Nan\c{c}ay during the frequent quiet periods
is almost free of electrical transients, even with a low trigger
threshold, thus offering excellent conditions for detecting EAS.

\subsection{Waveform observation}

\begin{figure}
\centering
  \vspace*{8pt}
\includegraphics[width=12cm]{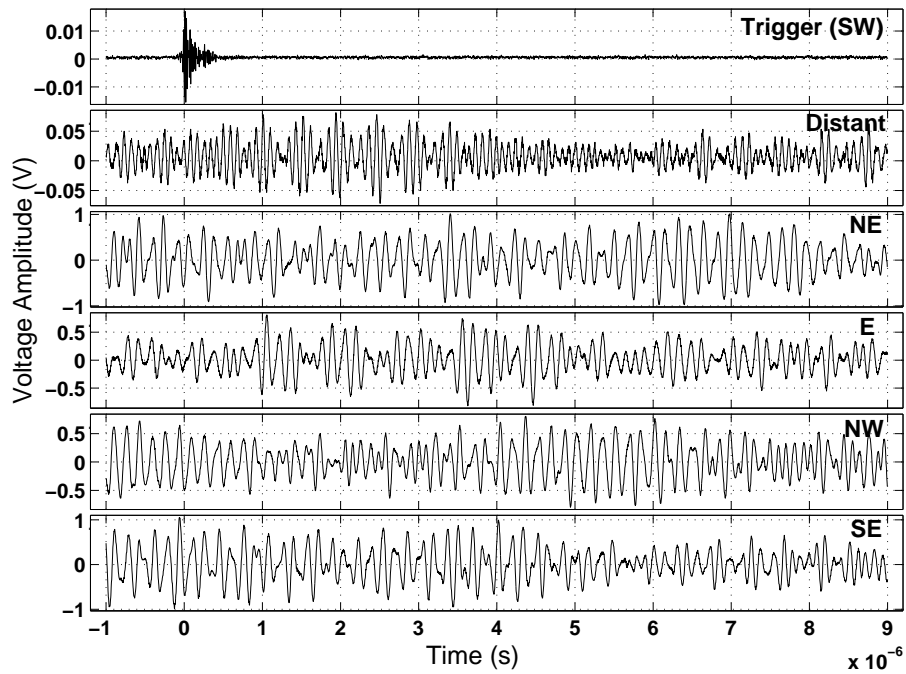}
\caption{ Waveforms from the six 
antennas for a typical event.  The trigger antenna is
filtered while the others are wide-band (see text).}
\label{fig:waveform_raw}
  \vspace*{12pt}
\end{figure}

In Fig.~\ref{fig:waveform_raw} the signal waveforms are shown for all
the antennas for a typical event.  The trigger pulse determines the
origin of the time scale. Due to the variable time-of-flight of the EM
signal (the contribution from cosmic rays corresponding to an
isotropic primary flux), the time involved in generating the trigger
and the effects of cable length, antenna pulses may precede the
trigger signal. Consequently, the oscilloscopes were set up to have a
1~$\mu$s pre-trigger time.  The trigger trace shows an oscillating
pulse due to the band-pass filter on this antenna as seen in section
\ref{subsec:filtering}. For all the other antennas, the expected
signal is hidden by the combined contributions from the radio
transmitters.

%
%
\section{Signal processing}
\label{sec:signal}

The first stage of the offline analysis is to find evidence for a
transient signal in the wide band channels and to determine its
associated time. Then the arrival direction of the incident electric
field is deduced if a coincidence is observed involving several
antennas. Based on shower properties, it is then possible to define
criteria for our device to select events from EAS. Finally in view of
a future step of the analysis, a possible method to recover the
waveform of the original signal is presented.

\subsection{Observation of coincidences}

\begin{figure}
\centering
  \vspace*{8pt}
\includegraphics[width=12cm]{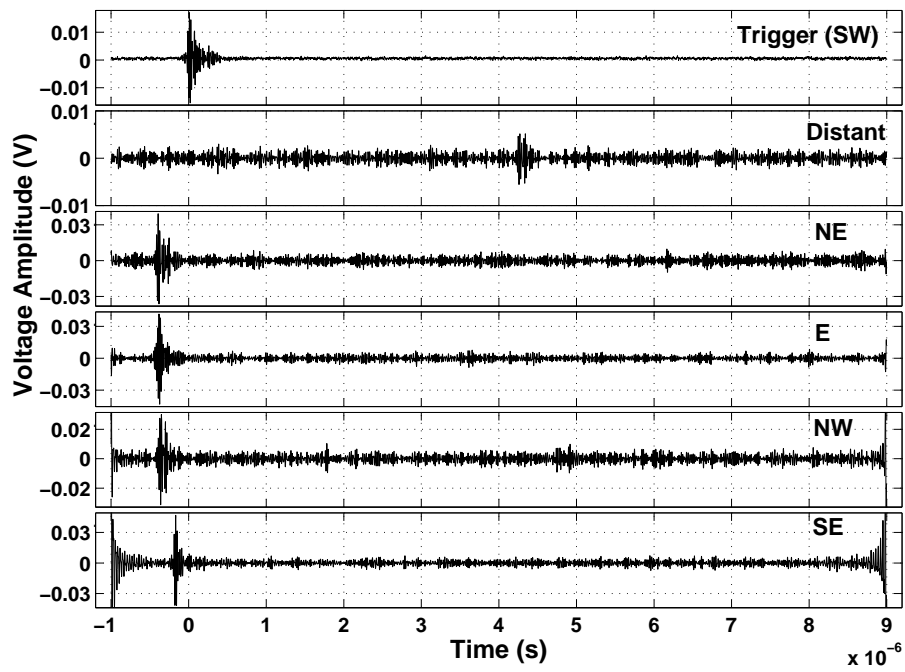}
\caption{Coincident transients in the 33--65~MHz
band obtained using numerical filtering  for the event presented in figure \ref{fig:waveform_raw}.}
\label{fig:waveform_filter}
  \vspace*{12pt}
\end{figure}

In order to reduce the broadcast contributions, we have filtered the
data numerically using the same frequency band as for the trigger
(33--65~MHz). Such a procedure is expected to enhance the presence of
any existing wide band signals. The effect of this band-pass filtering
is shown in Fig.~\ref{fig:waveform_filter} for the same event as in
Fig.~\ref{fig:waveform_raw}.  At each end of the time range, spurious
oscillations, coming from the Gibbs phenomenon, are further removed by
eliminating the two signal extremities (400 ns). In the remaining time
window, all antennas show a short
oscillating signal, which is characteristic of the result of band-pass
filtering on a transient pulse. The timing differences correspond to the
propagation times of the wavefront to the different antennas, and also
include electronics and cable delays. The signal detected on the
Distant antenna (0.8~km from the others) shows that transients can be
observed over large distances with our setup.

The major advantage of wide band data acquisition is the possibility to
filter the signal off-line in any desired frequency band.  In our
case, the low (1--5~MHz) band should be a valuable frequency window
(see section~\ref{sec:trigger_definition}) to provide additional
indications of the presence of broadband transients. However the
limited performance of the ADCs (8 bits) and the bandwidth of the
amplifiers used in this phase of the experiment have not yet permitted
us to perform such analysis. Currently, new 12 bit ADCs are being
installed, and after this upgrade it should be possible to investigate
this technique.

\subsection{Procedure for transient identification}
\label{subsec:sig_noise_windows}

In order to determine whether a transient pulse is present from a
particular antenna, we follow the filtering procedure just described.
A threshold is then set such that the signal voltage is required to
exceed the average noise, estimated on an event-by-event basis.  This
approach takes into account differences due to antenna location as
well as variations in radio conditions resulting from atmospheric
perturbations and human activities.

\begin{figure}
\begin{center}
  \vspace*{8pt}
\includegraphics[width=12cm]{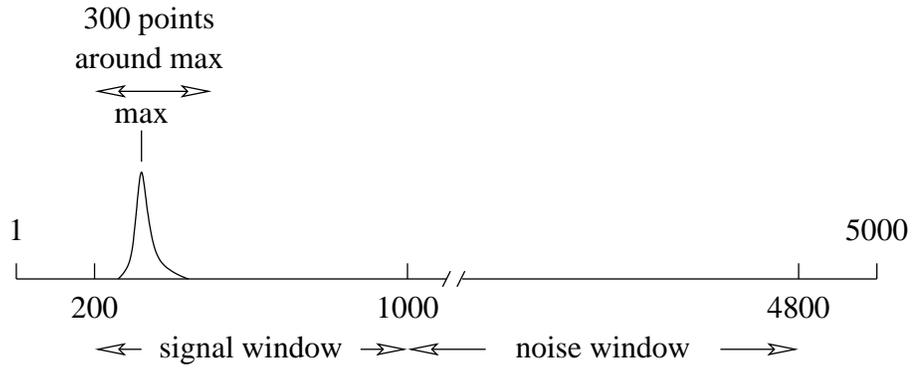}
\end{center}
\caption{Time windows used for the
 determination of signal and noise. The power average is calculated
 using the interval $[ i_\mathrm{max}-150:
 i_\mathrm{max}+150 ]$. When the maximum is close to the
 edge of the signal window, we use either the $[200:500]$ or
 the $[700:1000]$ interval.}
\label{fig:time_windows}
\end{figure}

As already mentioned, filtering the pulse generates an oscillating
pattern. In order to treat the positive and negative parts on an equal
footing, we use an antenna pulse power variable $P_i=v_i^2$ where $i$
is a time index with steps of 2 ns ($i=1$--5000) and $v_i$ is the
band-pass filtered pulse voltage at time~$i$.  The procedure is as
follows:
\begin{itemize}
\item the time range is divided into a signal window ($i=200$--1000) and a
  noise window ($i=1000$--4800) (see Fig.~\ref{fig:time_windows});
\item in the noise window, the noise average $\mu_n=\langle
  P\rangle_\mathrm{noise}$ and the standard deviation
  $\sigma_n=\sqrt{\langle (P-\mu_n)^2\rangle_\mathrm{noise}}$ are
  calculated;
\item a pulse transient is flagged when the power average
  $\mu_s=\langle P\rangle_\mathrm{signal}$ in a 300-point time range
  encompassing the $P_i$ maximum in the signal window (see
  Fig.~\ref{fig:time_windows}) is significantly greater than $\mu_n$,
  namely:
\[
\mu_s\ge \mu_n+k\frac{\sigma_n}{\sqrt{N_s}},
\]
  where $N_s$ is the number of points in the time window. The
  number $k$ has been adjusted empirically and $k=10$ has been found to
  provide unambiguous rejection of signals that are not pulse-like.
\end{itemize}

\subsection{Time tagging and triangulation}
\label{sec:triangulation}
The electromagnetic propagation time from one antenna to another 100~m
away is at most 330~ns (for a horizontal wave). This sets the scale
for the delays between two neighboring antennas in the network
(excluding the Distant antenna). Due to signal shrinking by Doppler
contraction, the pulse rise-time is expected to be small on this time
scale. The choice of a particular point on the waveform, e.g. maximum
or 10\% of the maximum, is therefore not a critical issue in obtaining
a reference time. Taking advantage of the correlation that exists
between the leading edge of the pulse and the maximum of the
oscillation envelope of the filtered signal (see
Fig.~\ref{fig:filtered_signal}), the pulse arrival time is taken to be
the point at which $P=v^2$ is maximum in the filtered band. The time
determination uncertainty due to the filtering is given by half the
pseudo-period of the filtered signal and is thus smaller than 20~ns.

The arrival times from the various antennas can then be used to
determine the incident direction of the electromagnetic wave. With the
hypothesis of a far source (at $10^{17}$~eV, vertical showers have
their maxima at an altitude of several km), a plane-wave
approximation is appropriate.  The plane's orientation is determined
by a fitting procedure and requires at least three coincident
antennas, since the following relation exists between arrival times
$t_j$, and antenna positions $(x_j,y_j,z_j)$:
\[
ct_j=\alpha x_j +\beta y_j+\gamma z_j+ct_0,
\]
where $(\alpha,\beta,\gamma)$ is the unit vector giving the EM field
incidence direction and $t_0$ the (unknown) time at which the
wavefront plane crosses the origin of the $(x,y,z)$ frame. Since, for
the flat Nan\c{c}ay site, the antennas are considered to lie in a
horizontal plane, we take $z_j=0$, and $\gamma$ cannot be determined
directly. The other parameters $\alpha, \beta$ and $t_0$ are
calculated using a least square fit to minimize the quadratic error
\[
\epsilon^2=\frac{1}{n_\mathrm{trig}}\sum_{j}
f_j\times(ct_j-\alpha x_j -\beta y_j -ct_0)^2
\]
where $f_j$ is 1 or 0 depending on whether a transient is flagged or
not on antenna $j$, and $n_\mathrm{trig}=\sum f_j$. When
$n_\mathrm{trig}\ge 4$ the value of $\epsilon$ gives the residual
error.  When coincidences occur that involve at least four antennas it
is possible to fit to a spherical wavefront instead of to a plane.
However, for a point source located at a distance $R$, much larger
than the antenna separation $\Delta$, the correction is of the order
of $\Delta^2/R$. This is larger than 10~m for $\Delta\sim100$~m only
if $R\le 1$~km so we consider the plane fit to be quite sufficient for
the air showers of interest.

The zenith ($\theta$) and azimuthal ($\varphi$) angles are determined
by inverting the relations
\[
\alpha=\sin\theta\cos\varphi,\quad\beta=\sin\theta\sin\varphi.
\]

\subsection{Event reconstruction}

\begin{figure}
\vspace*{8pt}
\begin{minipage}[t]{\textwidth}
   \centering
   \includegraphics[width=9.cm]{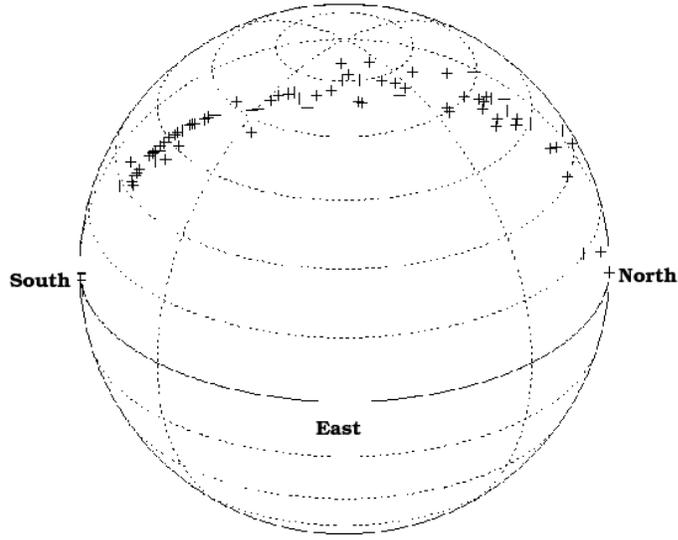}
   \vspace*{8pt}   
\end{minipage}
\begin{minipage}[t]{\textwidth}
\centering
\includegraphics[width=9.cm]{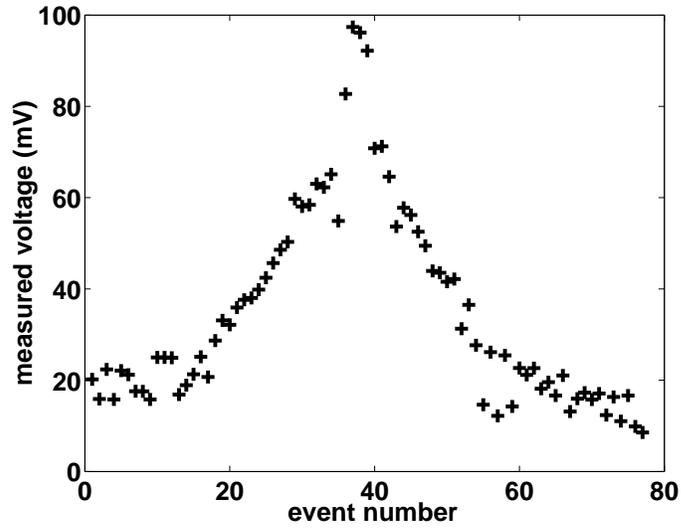}
\end{minipage}
\caption{Up: Location in the sky of the mobile source. Down:
Maximum voltage recorded on one of the antennas during this series of
triggers.}
\label{fig:avion}
  \vspace*{12pt}
\end{figure}

The capability of the device to reconstruct signal directions is
illustrated below using a somewhat atypical data set recorded by
CODALEMA consisting of nearly a hundred successive triggers that
occurred within a 2~minute period. This high trigger rate would
suggest that those events are from interference of human or
atmospheric origin.  In Fig.~\ref{fig:avion} the wavefront direction
for each event is plotted on a sky map.
The sequence of points form a trajectory which would correspond to
that of a single emitting source moving from north to south above the
detector array.

The maximum measured voltage on one of the antennas for this event
sequence is plotted to the right in Fig.~\ref{fig:avion}. As expected
for a single source moving on a more or less linear trajectory above
the detector, the maximum measured voltage increases as the source
approaches, then decreases when it moves away.  The interpretation of
these observations is that the source is an aircraft or
possibly a satellite passing over Nan\c{c}ay.

\begin{figure}
\centering
  \vspace*{8pt}
\includegraphics[width=10.cm, height=10.cm]{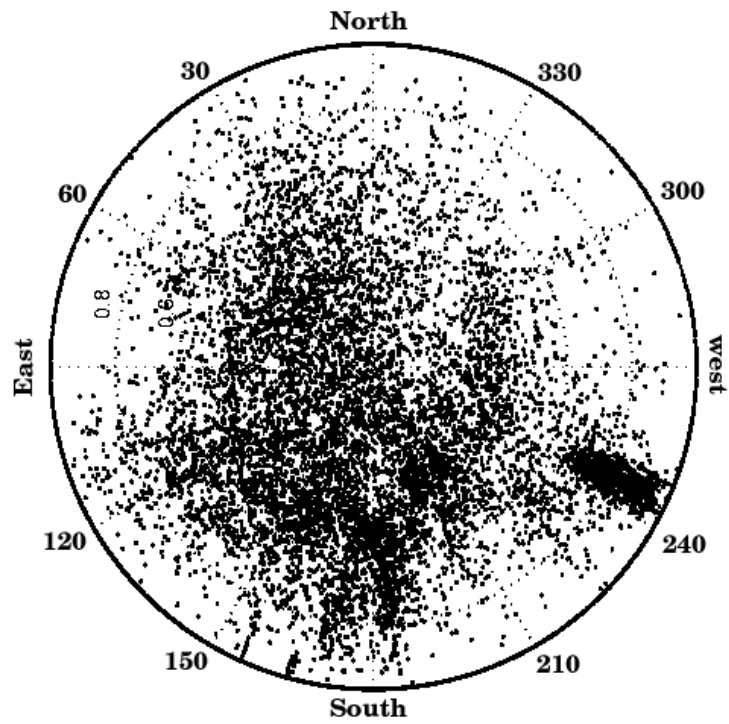}
\caption{Projection on the ground plane of the directions of $11\times
  10^3$ reconstructed events among the $50\times 10^3$ recorded by
  CODALEMA between March 2003 and June 2004. An accumulation of events close to the horizon is visible 
from the West-South-West dues to human activities in the vicinity of the station.}
\label{fig:lob-antenna}
  \vspace*{12pt}
\end{figure}

Fig.~\ref{fig:lob-antenna} shows the directional distribution of the
11000 reconstructed events resulting from a year of data taking.

A substantial number of events had a below-horizon direction and are not
plotted in the figure. With the help of a spherical fit, most of these
were found to originate from electrical devices located at various
points on the Nan\c{c}ay observatory site. For the reconstructed
events in the plot, their distribution indicates that the detector
array has higher sensitivity for signals coming from the south. This
is not surprising since the antennas are inclined to the south
(tilting of $25^\circ$) to facilitate observations of the sun and Jupiter.

\subsection{Event selection}

Due to the steep falloff of the cosmic ray flux with energy, most of
signals that produce a trigger should correspond to 10$^{17}$~eV EAS
(see section~\ref{sec:1017}). Based on the properties of these
showers, any such events must fulfill several conditions.  At this
energy, showers with small impact parameters, i.e.\ whose axis is
close to or inside the square formed by the four DAM antennas, are
expected to provide the largest signal amplitudes. As a consequence,
the presence of tagged signals on the four antennas constitutes the
first experimental criterion (referred to as ``square'') required to
select events from EAS.  In addition, the distant antenna would only
be tagged by very energetic and very unlikely events or by
interference caused by kilometer-scale RF perturbations. Consequently,
the absence of tagged signals from this antenna gives the second
criterion (referred to as ``not distant'') that must be
fulfilled. Finally, since at small impact parameters the electric
field is maximum for vertical showers, we require as an additional
condition (referred as ``$\theta\le 45^{\circ}$''), that the zenith
angle be less than $45^\circ$. This last criterion also helps us to
remove interference that tends to come from the ground level or low
elevations.

\begin{figure}
\begin{center}
\includegraphics[height=8cm]{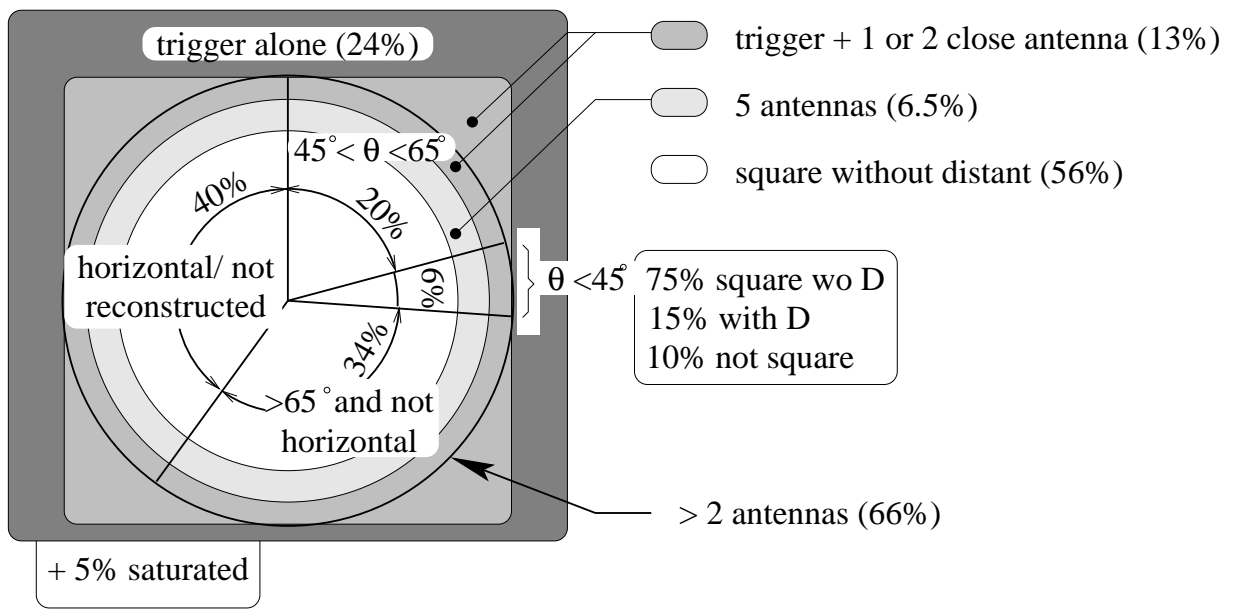}
\end{center}
\caption{Sketch of the selection cuts used to find the EAS candidates
and the corresponding distributions of the event topologies.}
\label{fig:distribution_2mV}
\end{figure}

The best performance was reached using the restricted band
(24--82~MHz) with a low trigger threshold (2~mV). The necessary
conditions for satisfactory operation were satisfied only during a few
nights (see section~\ref{sec:digit}).  A careful study of these runs
was undertaken and the events were classified using several
criteria. The resulting distributions are presented in
Fig.~\ref{fig:distribution_2mV}. It turns out that only 3\% of the 3.$10^{3}$
events meet the combined criterion ``square \& not Distant \&
$\theta\le 45^{\circ}$''.

\begin{figure}
\begin{center}
\includegraphics[height=10cm]{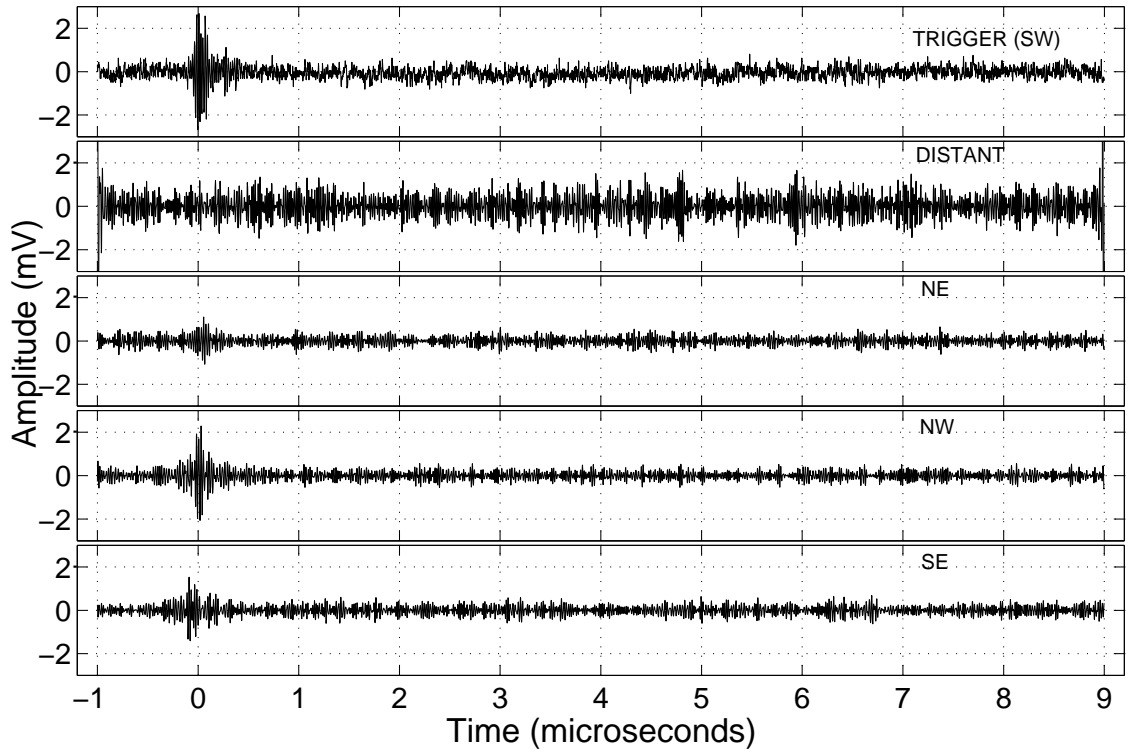}
\end{center}
\caption{Air shower candidate signals after filtering in the 33--65~MHz band for the event fulfilling all our criteria. The reconstructed
direction is $\theta = 41 ^\circ$ and  $\phi = 161 ^\circ$.}
\label{fig:candidateas}
\end{figure}

The next step in the selection process was to remove any event
containing something ``suspicious'' such as multiple transient pulses
within the 10~$\mu$s data record, long duration transient signals
($>1~\mu$s) or pulses coming from almost the same location in the sky
occurring each day around the same time. 
This was done on an event-by-event basis.  Finally
after this very severe scrutiny of candidates, only one event from
this data subset appears as a possible EAS candidate (see
Fig.~\ref{fig:candidateas}). This result would seem to be compatible
with the cosmic ray flux for energies $\ge 10^{17}$~eV which is
$0.2/$km$^2/$hour$/$sr; taking the solid angle corresponding to
$\theta\le 45^{\circ}$ and a surface of collection in the range
$0.01-0.1$~km$^2$ leads to counting rates from one per 10 days up to
one per day.  Only further data taking under our best sensitivity
conditions together with coincident information from particle
detectors will provide us with a better understanding
of the acceptance of our device and lead to the appropriate criteria
for EAS identification with a self-triggering radio setup.

\subsection{Waveform restoration}

Waveform information cannot be recovered from the frequency-filtered
data, but knowledge of the waveform is mandatory for a complete
physical analysis which could contribute to the characterization of
the primary cosmic ray. The present section is devoted to discussing
possible processing tools able to recover the pulse shape from the raw
data.

\begin{figure}
 \centering
   \vspace*{8pt}
\includegraphics[width=10cm]{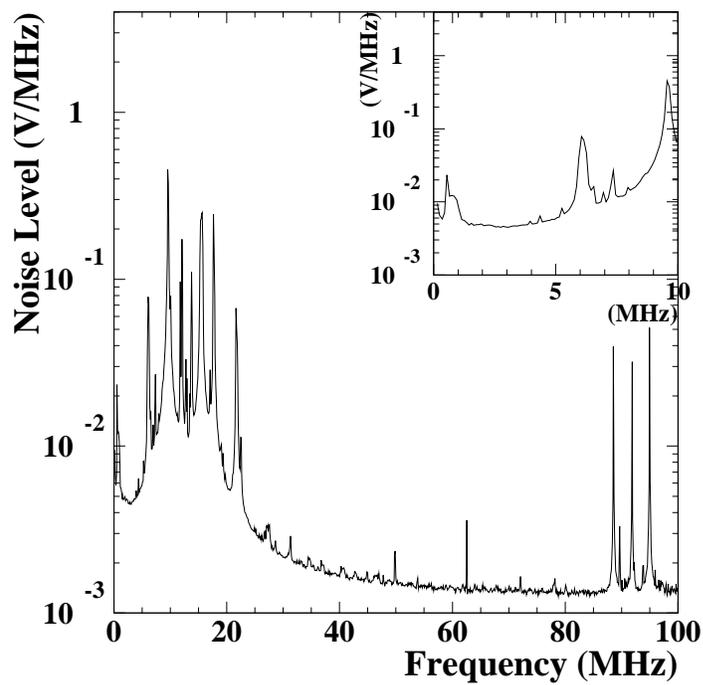}
  \caption{Mean direct Fourier transform of 900 10~$\mu$s noise
  snapshots measured at Nan\c{c}ay as a function of the frequency. The
  inset presents a zoom of the curve for frequencies below
  10MHz. The leakage phenomenon is clearly visible in the shape of the
  spectrum (to be compared with the power spectral density presented
  in Fig.~\ref{fig:mean_spectre_radio_Nancay}).}
  \label{fig:leakage}
\end{figure}

A typical signal and the radio noise spectrum are shown in
Fig.~\ref{fig:waveform_raw} and
Fig.~\ref{fig:mean_spectre_radio_Nancay} respectively.
In principle, after transient detection, signal recovery could be
achieved by eliminating the various transmitter lines from the
frequency spectrum.  A simple way to do this is to make a direct
subtraction of a noise spectrum derived from the ``noise window'' from
the equivalent ``noise+signal'' spectrum obtained from the ``signal
window'' (see section~\ref{subsec:sig_noise_windows}). It was found
that this gives poor results due to the strong variability of AM
transmitter amplitudes in the 6--25~MHz frequency range. Only signals
in the FM band show amplitude stability during the 10 $\mu$s time
window.  A second difficulty with the strong AM frequency components
inherent to the time-bounded nature of data is the leakage phenomenon~\cite{numrecip} illustrated in Fig.~\ref{fig:leakage}.
It can be
considerably reduced by performing amplitude limiting on the strongest
components above an appropriate threshold~\cite{casa-mia}. This
threshold value is not constant, because reception conditions can vary
from one event to another.  Taking advantage of the great stability of
FM transmitter amplitudes in the 10~$\mu$s record, the amplitude
limiting threshold was set at a quarter of the peak value in the FM
band. This simple shrinkage method gives excellent results for
reducing the leakage phenomenon.

However, though a strong transient could possibly be made visible with
this technique, its shape is always modified by AM and FM transmitter
residuals. In order to clean the signal, a subsequent processing step
is necessary. One possibility is to extract the waveform
characteristics from the measured frequency spectrum assuming a simple
analytical expression for the pulse shape.
 
As noted in section~\ref{sec:motivation}, the analytical expression of
the pulse given by Eq.~\ref{eq:waveform} can be simplified for this
purpose. For vertical air showers with small impact parameters, those
which give rise to the largest radio signals, $\tau_2$ is expected to
be much greater than $\tau_1$. Furthermore, since the frequency band
24--82~MHz lies above $1/(2\pi\tau_1)$, the $\tau_2$ term gives only a
negligible contribution to the Fourier transform in this band and it
is neglected. The
modulus of the Fourier transform vs frequency $\nu$ then can be
written
\begin{equation}\label{eq:modulus}
|S(\nu)|=\frac{ A e^2 \tau_1 }{ 2 }
\frac{ 1 }{(1 + 4\pi^2\nu^2\tau_1^2)^{3/2}}
\end{equation}
A least squares fit of the spectrum using this expression provides
values for $\tau_1$ (in ns) and for the amplitude factor A (in volts),
thus determining the shape of $S(\nu)$ at all frequencies.

\begin{figure}
  \vspace*{8pt}
\begin{minipage}[t]{.45\textwidth}
   \includegraphics[width=6.5cm]{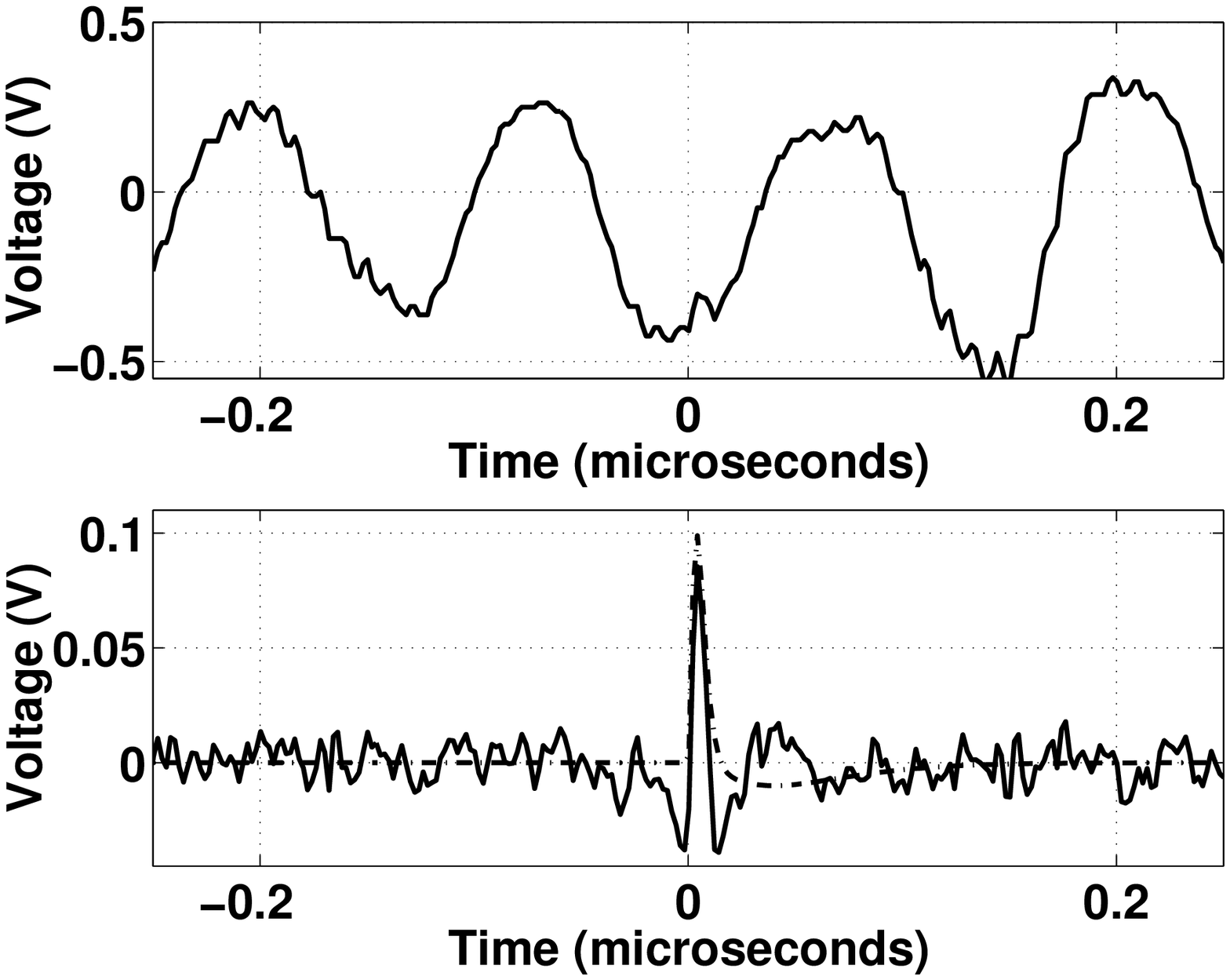}
   \vspace*{8pt}
\end{minipage}
\hfill
\begin{minipage}[t]{.45\textwidth}
 \includegraphics[width=6cm]{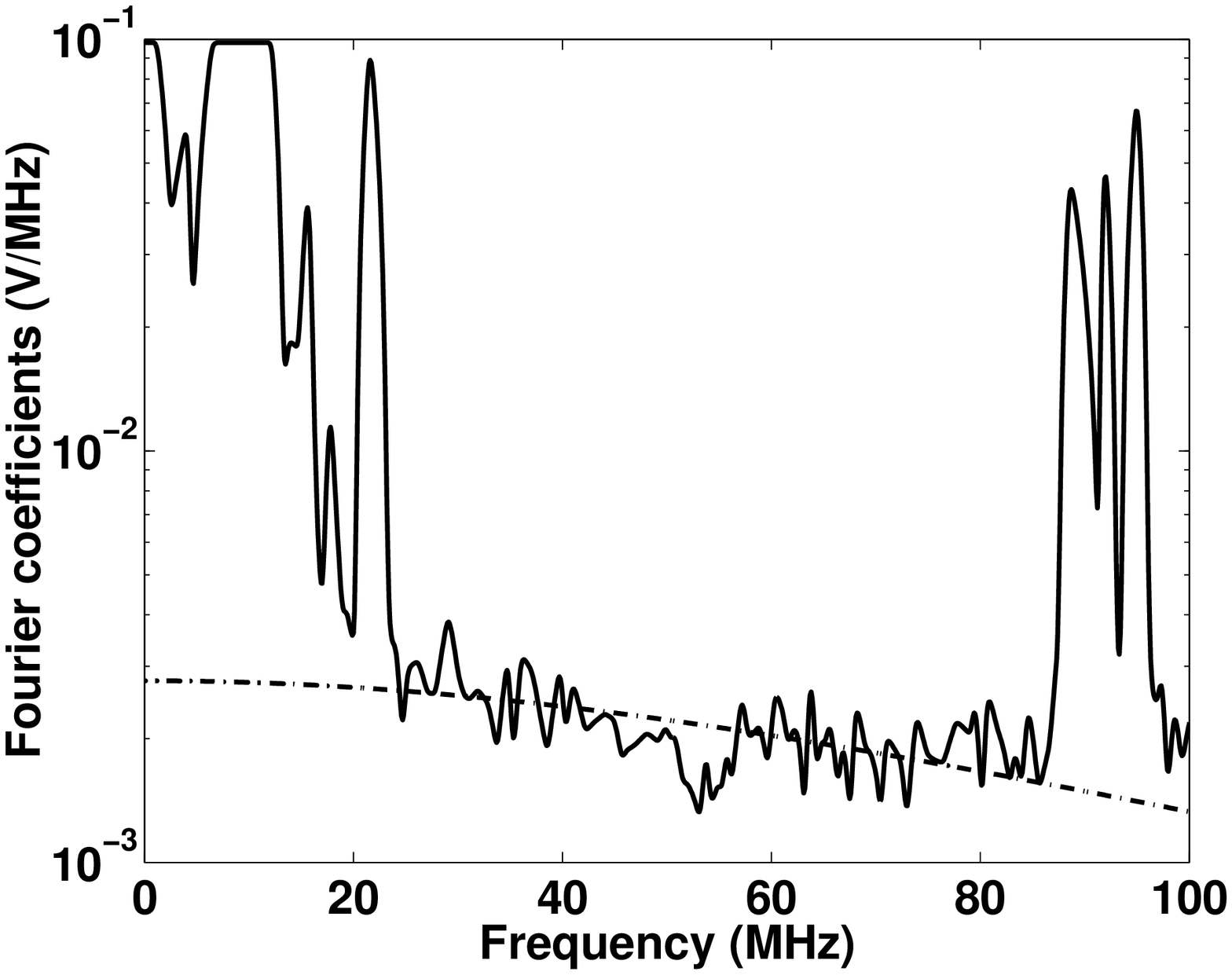}
  \vspace*{8pt} 
\end{minipage}
\caption{Illustration of signal shape recovery by spectrum
fitting. Top left: simulated signal, made up of a typical recorded
noise sequence and a pulse added at 0~ns. Bottom left: reconstructed
signal in the time domain. The pulse is made visible, though its shape
is slightly modified with respect to the simulated one (dash-dotted
line). Right: envelope of the Fourier transform modulus (solid line)
together with its fit in the 24--82~MHz band (dash-dotted line) after
amplitude limiting of the coefficients in the AM transmitter
band.}
\label{fig:fit_example}
\end{figure}

As an example, Fig.~\ref{fig:fit_example} illustrates the whole
process applied to a simulated signal made up of a typical noise
record from one antenna, to which has been added a transient beginning
at 0~ns whose shape is given by formula~\ref{eq:waveform}.  For the
full-band configuration the signal is recovered reasonably well for a
minimum pulse amplitude of 50~mV and $\tau_1\ge 2$~ns.  At the moment,
the limitation in using this technique experimentally is set
principally by our poor ADC resolution.  Considering electronic and
antenna gains, this sets a lower limit on the peak electric field
value of 1.4~mV/m. The method can be used on data from the restricted
band for $A\ge$5~mV and $\tau_1=2-8$~ns.

This process was applied to the EAS candidate signal of
Fig.~\ref{fig:candidateas}, and gave peak amplitudes A and rise-times
$\tau_1$ of 16~mV and 4.0~ns, 16~mV and 7.7~ns, and 13~mV and 3.6~ns
for the NE, NW and SE antennas respectively. Considering the overall
gain factor of the reception chain, the electric field that triggered
had a peak value of about 0.4~mV/m. 


\section{Coincidence measurements}
\label{sec:particules}
\begin{figure}
\begin{center}
\includegraphics[width=8.3cm]{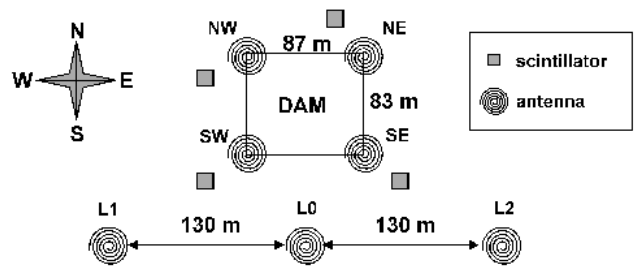}
\end{center}
\caption{CODALEMA setup for the second phase. The particle detector
acts as a trigger with a fourfold coincidence requirement.}
\label{fig:setup2}
\end{figure}

In order to investigate possible correlations between measured radio transients and air showers, four stations of
particle detectors acting as a trigger have been added.
This new setup, including a new
antenna configuration, is shown in Fig.~\ref{fig:setup2}.  It uses
seven log-periodic antennas with their electronics (see section \ref{sec:digit}).
To get enough sensitivity
with our ADCs, all the antenna signals are band pass filtered (24-82 MHz). In
this configuration, all the antennas are treated identically.

\subsection{Particle detectors}
The trigger corresponds to a fourfold coincidence within 600~ns from
the particle detectors. These were originally designed as a prototype
detector element for the Auger array, consisting of four plastic
scintillator modules~\cite{boratav}. Each 2.3~m$^2$ module (station)
has two layers of acrylic scintillator, read out by a single
photomultiplier placed at the center of each sheet. The
photomultipliers have copper housings providing electromagnetic
shielding. The signals from the upper layers of the four stations are
digitized (8-bit ADC, 100~MHz sampling frequency, 10~${\mu}$s
recording time).

A station produces a signal when a coincidence between the two layers
is obtained within a 60~ns time interval. This results in a counting
rate of around 200~Hz. The rate of the fourfold coincidence of the
four stations is about 0.7 events per minute, corresponding almost
entirely to air shower events.  The four stations are located close to
the corners of the DAM array, and in this configuration the particle
detectors delimit an active area of roughly~$7\times 10^3$~m$^2$.
Using arrival times from the digitized PMT signals, it is possible to
determine the direction of the shower by triangulation with a plane
fit (same procedure as section \ref{sec:triangulation}).
  From the arrival direction distribution, a value of $16\times
10^3$~m$^2\times$sr is obtained for the acceptance, which corresponds
to an energy threshold of about $1\times 10^{15}$~eV.

\subsection{Event analysis}
For each fourfold coincidence from the particle detectors, the seven
antenna signals are recorded. Due to the relatively low energy
threshold of the trigger system, only a small fraction of these air shower events is
expected to be accompanied by significant radio signals.

To identify these events, an offline analysis (see section \ref{sec:signal})
is made. When at least 3 antennas are flagged it becomes possible to apply a 
triangulation procedure and the
event is declared to be a radio candidate if the arrival 
direction obtained is above the horizon.

The counting rates are the following: 1 event per hour for a single 
antenna-trigger coincidence, 1 event every 2 hours for a three-fold antennas-trigger 
coincidence. This indicates that the energy threshold for radio detection
is substantially higher than that of the particle array and also that
the antennas do not regularly pick up electrical signals related to
particle detector activity.  Of course, in the 2~$\mu$s window where
the search is conducted, a radio transient can occur which is not
associated with the air shower. Being uncorrelated, such events should
have a uniform arrival time distribution. The next step in the
analysis was therefore to study the arrival time distribution of radio
signals.

\begin{figure}
\begin{center}
\includegraphics[width=8.3cm]{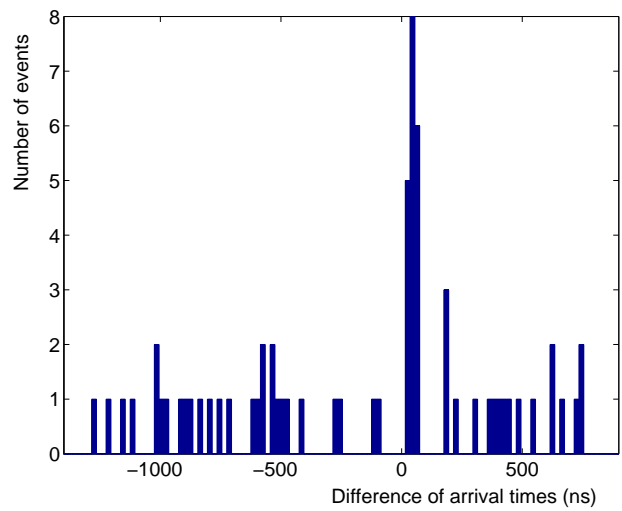}
\end{center}
\caption{Time delay between the radio plane front and the particle
  plane front}
\label{fig:deltat.eps}
\end{figure}

For radio events that originate from air showers (and possibly from
scintillator activity), the radio-particle correlation should manifest
itself as a peak in the antenna arrival time distribution, the time
reference being furnished by the particle trigger. The width of the
raw distribution is primarily determined by the shower arrival
direction variation and thus depends on the antenna locations with
respect to the particle detectors. For a vertical shower, detectors
and antennas will be hit essentially at the same time, neglecting
small delays related to the curvature and width of the shower front or
small differences between the electromagnetic wave and shower particle
propagation times. On the other hand, in the extreme case of a
horizontal event coming from the SW direction, the L1 antenna (see
Fig.~\ref{fig:setup2}) can receive a signal 0.7~$\mu$s before the
scintillator located at the NE corner of the DAM array. 
The time of passage of the radio wavefront through a reference point is 
compared to the particle front time extracted from the scintillator signals.
 This time delay distribution is shown in
Fig.~\ref{fig:deltat.eps}. The data correspond to 59.9 acquisition
days and 70 antenna events.
A very sharp peak (a few tens of nanoseconds) is obtained showing an
unambiguous correlation between certain radio events and the particle
triggers. This peak is not exactly centered on zero, the mean value
being about 40~ns. The systematic error on these time differences, due
to inaccuracies in determining signal times has been estimated to be around 20~ns. 
The delay between the electric field and particle
shower times could be measured with our apparatus, though more
thorough studies and higher statistics will be necessary. As
expected, there is also a uniform distribution in the 2~$\mu$s window
corresponding to accidentals.

A similar arrival time correlation would also have been
obtained if the signal on each antenna had been directly induced by a
nearby particle detector. However in the present configuration, such
an effect can be excluded since three of the seven antennas, which
detect signals comparable to those from the other antennas, are not
located close to particle detectors.  Moreover, it has been verified 
that no correlation exists between high photomultiplier signal amplitude 
and the presence of antenna signals.

\begin{figure}
\begin{center}
\includegraphics[height=8.3cm,angle=90]{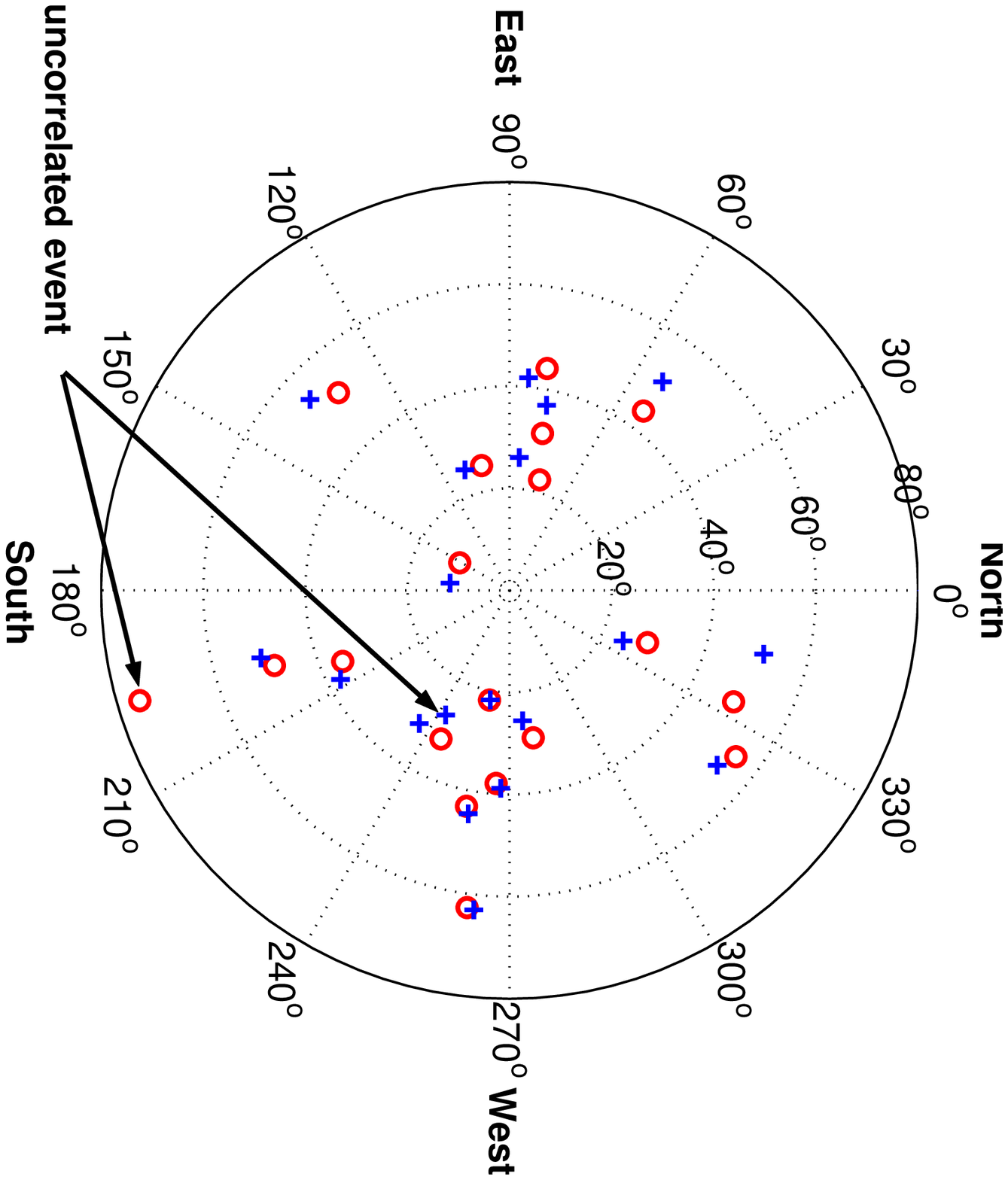}
\end{center}
\caption{Event arrival directions. The circles indicate directions
reconstructed from antenna signals whereas the crosses correspond to
directions given by the scintillators. Except for the marked event, each
circle is associated with the nearest cross.}
\label{fig: Event arrival direction}
\end{figure}

Finally, if the time-correlated events correspond to extensive air
showers, they must also have correlated arrival directions.
Fig.~\ref{fig: Event arrival direction} shows, for the 19 events
located in the main peak of Fig.~\ref{fig:deltat.eps}, the arrival
directions reconstructed from both scintillator and antenna
data. Except for one event, each antenna direction is associated with
the nearest scintillator value. The distribution of the angle between
the two reconstructed directions is as expected, \textit{i.e.}, a
gaussian  centered on zero multiplied by a sine function
coming from the solid angle factor. The standard deviation of the corresponding
gaussian is about 4 degrees. The one event
with a much bigger angular difference is certainly an accidental. The arrival
 direction given by the antennas is close to the
horizon, which is typical of events from radio interference due to
human activity.  Moreover, the presence of one chance event in the
peak is compatible with the observed uniform distribution in the
2~$\mu$s window.

These results strongly support the claim that electric field
transients generated by extensive air showers have been measured with
CODALEMA, and that the incident direction of the primary cosmic ray
can be reconstructed from the arrival times of radio signals.

For the 18 EAS events of Fig.~\ref{fig: Event arrival
direction} the antenna multiplicity varies from 3 to 7 and the electric field
amplitude of the filtered signals goes up to $1.2$~mV$/$m. This
signal level corresponds to typical values expected for air shower
energies in the $ 10^{17}$~eV range~\cite{Allan}.

In order to obtain a rough estimate for the energy threshold of the
present experimental setup, we suppose that the acceptance is the same
for the two types of detectors (\textit{i.e.} $16\times
10^3$~m$^2\times$sr). The observed event rate then leads to an
approximate energy threshold of $5\times 10^{16}$~eV, using the $\frac{1}{E^2}$ power 
law of the integral cosmic ray flux.

\section{Conclusion}
\label{sec:conclusion}

The CODALEMA experiment at the Nan\c{c}ay Radio Observatory in
central France has shown that this site offers a very good
electromagnetic environment for the observation of transient signals
with broadband frequency characteristics.
Investigation covering a wide frequency span conjugated
with numerical filtering and event-by-event noise estimates are the
main features which have been implemented in the transient
identification procedure. 

 Convincing evidence has been given for the observation of radio
signals associated with extensive air showers by CODALEMA using the coincidence measurements. 
The originality of this method lies in the possibility of our antenna array to work in 
a self triggering mode. In particular, we
have shown that our antenna array and analysis methods can be used
successfully to produce time values and topological information to
reconstruct the arrival direction and event waveform. 
The present results clearly demonstrate the interest of a complete
re-investigation of the radio detection method proposed by Aska'ryan
in the 1960's when the performance of the computing and electronics
equipment then available was insufficient to identify clearly
transient signals from EAS and to discriminate against background.

Improvements in the experimental setup that are in progress include,
firstly, additional scintillators to be installed at Nan\c{c}ay to make
possible the determination of the shower energy and core position. Two
antennas will also be added on each side of the existing W-E line,
thereby increasing its length to 600~m providing better sampling of
the radio shower signal extension.
 In a second step, the sensitivity
of the array will be increased by the use of 
12-bit encoding. This will allow us to record the
full 1-100 MHz frequency band and thus to infer
shower parameters from the full signal shape.
Shower
parameters could then be inferred from the signal shape. A dedicated
processing tool is being built both for detection and waveform
recovery.

In a subsequent upgrade, it is planned to increase the antenna area by
the installation of dipole antennas equipped with active front-end
electronics. This front-end will use a dedicated ASIC amplifier (gain
35~dB, 1~MHz~-~200~MHz bandwidth, $1.3~\mathrm{nV/\sqrt{Hz}}$)
currently under test. Furthermore, future antennas will be
self-triggered and self-time-tagged, coincident events being
recognized offline.

The latter point is part of current investigations concerning the feasibility
of adding radio detection techniques to an existing surface detector
such as the Pierre Auger Observatory. The radio signals should provide
complementary information about the longitudinal development of the
shower, as well as the ability to lower the energy threshold
(depending on the antenna array extension).

Considering that radio detection is expected to offer several specific
advantages such as a high duty cycle and a large sensitive volume at
moderate cost, this method opens up new prospects for supplementing existing large
hybrid detection arrays such as the Pierre Auger Observatory in view of
elucidating the enigma surrounding the origin of the highest energy cosmic
rays.


\clearpage


\begin{table}[h]
\centering
\caption{Electronic sensitivity for the different operating configurations
used in the experiment. The last row gives the antenna-related electric field
sensitivity.}
  \vspace*{8pt}

\begin{tabular}{|c |c |c |c|}
\hline
             &  Trigger Band     & Full Band       & Restricted band  \\
             &  33--65 MHz& 1--100 MHz & 24--82 MHz \\
\hline
ADC dynamic      & $\pm $ 8 mV         & $\pm$ 1200  mV    & $\pm $80 mV \\
\hline
ADC resolution &  0.0625 mV/bit         & 9.375 mV/bit           &
             0.625 mV/bit     \\
\hline
Sensitivity at 50MHz & 2 $\mu$V/m/bit  & 300 $\mu$V/m/bit &  20
             $\mu$V/m/bit\\
\hline
\end{tabular}
\label{tab: sensitivity}
  \vspace*{12pt}
\end{table}

\hfill\eject
\clearpage
\clearpage
\clearpage
\clearpage

\clearpage
\clearpage
\clearpage
\clearpage

\clearpage

\clearpage

\clearpage

\clearpage

\clearpage

\clearpage

\clearpage

\clearpage
\clearpage
\clearpage

\clearpage

\clearpage


\begin{thebibliography}{00}

\bibitem{auger} Auger Collaboration, Pierre Auger Project Design Report
(2nd Edition, Nov. 1996, rev. Mar. 1997) Fermilab, 1997 (available from
http://www.auger.org); J.W. Cronin, Rev. Mod. Phys. 71, S165 (1999).

\bibitem{Ask62} G.A. Askar'yan, Soviet Physics J.E.T.P., 14 (1962)
  441.

\bibitem{sigl} P. Bhattacharjee, G. Sigl, Phys. Rep. 327 (2000) 109.

\bibitem{Allan} H.R. Allan, in: Progress in elementary particle and
cosmic ray physics, ed. by J.G. Wilson and S.A. Wouthuysen (North
Holland, 1971), p. 169.

\bibitem{agasa} N. Hayashida et al., Phys. Rev. Lett. 73 (1994)
3491. M. Takeda et al., Phys. Rev . Lett. 81 (1998) 1163.

\bibitem{fly} D.J. Bird et al., Phys. Rev. Lett. 71 (1993) 3401, and
Astrophys. J. 441 (1995) 144.


\bibitem{casa-mia} K. Green, J.L. Rosner, D.A. Suprun, J.F. Wilkerson,
Nucl. Instrum. Meth. A498 (2003) 256.

\bibitem{Lopez}  A. Horneffer, H. Falcke, A. Haungs, K.H. Kampert,
 G.W. Kant, H. Schieler, proceedings of the 28$^{th}$ International Cosmic
 Ray Conferences (ICRC 2003), Tsukuba, Japan, 31 Jul - 7 Aug 2003 (to be
 published); H. Falcke, P. Gorham, Astropart. Phys. 19 (2003) 477.

\bibitem{Lofar} http://www.lofar.org

\bibitem{cascade-grande} A. Badea \textit{et al.}, to appear in the
  proceedings of the 5$^{th}$ Cosmic Ray International Seminar: GZK
  and Surroundings (CRIS 2004, Catania, Italy, 2004),
  astro-ph/0409319.

\bibitem{rav04} O.Ravel, R.Dallier, L.Denis, T.Gousset, F.Haddad,
P.Lautridou, A.Lecacheux, E.Morteau, C.Rosolen, C.Roy, Proceedings of
the 8th Pisa Meeting on Advanced Detectors ``Frontier Detectors for
Frontier Physics'', \textit{Nucl. Instr. Meth.} \textbf{A518} (2004)
213.

\bibitem{dal03} R.Dallier, L.Denis, T.Gousset, F.Haddad, P.Lautridou,
A.Lecacheux, E.Morteau, O.Ravel, C.Rosolen, C.Roy, \textit{SF2A 2003
Scientific Highlights}, ed F. Combes \textit{et al.} (EDP Sciences,
2003).

\bibitem{dal04} A. Belletoile, D.Ardouin, D. Charrier, R.Dallier,
L.Denis, P. Eschstruth, T.Gousset, F.Haddad, J. Lamblin, P.Lautridou,
A.Lecacheux, D. Monnier-Ragaigne, A. Rahmani, O.Ravel, \textit{SF2A
2004 Scientific Highlights}, ed F. Combes \textit{et al.} (EDP
Sciences, 2004), astro-ph/0409039 (2004).

\bibitem{slac} D. Saltzberg, P. Gorham, D. Walz, C. Field, R. Iverson,
 A. Odian, G. Resch, P. Schoessow, D. Williams, Phys. Rev. Lett. 86
 (2001) 2802.

\bibitem{rice} J.A. Adams et al., Proceedings of the 4th Tropical
 Workshop on Particle Physics and Cosmology: Neutrinos, Flavor Physics
 and Precision Cosmology, Cairns, Queensland, Australia, 9-13 Jun
 2003, AIP Conf.Proc.689:3--15,2003.

\bibitem{moon} P.W. Gorham, K.M. Liewer, C.J. Naudet, Proceedings
26$^{th}$ International Cosmic Ray Conference, Salt Lake City, Utah,
17--25 Aug 1999, vol. 2* 479--482, astro-ph/9906504.

\bibitem{kandl} F.D. Kahn, I. Lerche, Proc. Roy. Soc. A 289 (1966)
  206.

\bibitem{huege} T. Huege, H. Falcke, Astronomy \& Astrophysics, 412
 (2003) 19; T. Huege, H. Falcke, astro-ph/0409223.

\bibitem{jackson} J.D. Jackson, \textit{``Classical
  Electrodynamics''} (Wiley, New York, 1975).

\bibitem{gousset} T. Gousset, O. Ravel, and C. Roy, Astropart.
Phys., 22 (2004) 103.

\bibitem{ardouin} D. Ardouin, A. Belletoile, D. Charrier, R. Dallier,
L. Denis, P. Eschstruth, T. Gousset, F. Haddad, J. Lamblin,
P. Lautridou, A. Lecacheux, D. Monnier-Ragaigne, A. Rahmani, O. Ravel,
Proceedings of the 19th European Cosmic Ray Symposium, Florence, 2004,
astro-ph/0412211.

\bibitem{suprun} D.A. Suprun, P.W. Gorham and J.L. Rosner,
  Astropart. Phys. 20 (2003) 157.

\bibitem{RDN} http://www.obs-nancay.fr/html\_fr/decametr.htm 


\bibitem{GAP} J. Lamblin , O. Ravel and C. Medina, internal report SUBATECH 03/2005.

\bibitem{kraus} J.D. Kraus, \textit{``Antennas''} (McGrawHill, 1988).

\bibitem{numrecip} W.H. Press, S.A. J.D. Teukolsky, W.T. Vetterling,
B.P. Flannery, \textit{``Numerical recipes in C''} (Cambridge University Press, 1992).

\bibitem{boratav} M. Boratav, J.W. Cronin, B. Dudelzak, P. Eschstruth,
  P. Roy, V. Sahakian and Z. Strachman, The AUGER Project: First
  Results from the Orsay Prototype Station, Proceedings of the 24th
  ICRC, Rome, 954,(1995).


\end{thebibliography}
\end{document}